\documentclass[11pt,a4paper]{article}
\usepackage[left=2.35cm,top=3cm,bottom=3cm,right=2.3cm]{geometry}

\usepackage{amsmath}
\usepackage{amssymb}
\usepackage{color}
\usepackage{cite}
\usepackage{graphicx}
\usepackage{subcaption}
\usepackage{slashed}
\usepackage{multirow}
\usepackage{upgreek}
\allowdisplaybreaks

\usepackage[font=small]{caption}

\usepackage[colorlinks=true
,urlcolor=blue
,anchorcolor=blue
,citecolor=blue
,filecolor=blue
,linkcolor=blue
,menucolor=blue
,linktocpage=true
,pdfproducer=medialab
]{hyperref}

\usepackage{listings}

\newcommand{\mcmule}{{\sc McMule}}

\newcommand{\cM}{\mathcal{M}}
\newcommand{\cA}{\mathcal{A}}

\renewcommand{\Re}{\mathrm{Re}}

\newcommand{\szet}{\sigma_0}
\newcommand{\soel}{\sigma^{(1)}_e}
\newcommand{\soml}{\sigma^{(1)}_\mu}
\newcommand{\soxl}{\sigma^{(1)}_x\,\Bigl\{^+_-}
\newcommand{\sopl}{\sigma^{(1)}_p}
\newcommand{\solf}{\sigma^{(1)}_\Pi}

\newcommand{\stel}{\sigma^{(2)}_e}
\newcommand{\stml}{\sigma^{(2)}_\mu}
\newcommand{\stxl}{\sigma^{(2)}_x\,\Bigl\{^+_-}
\newcommand{\stpl}{\sigma^{(2)}_p}
\newcommand{\stef}{\sigma^{(2)}_{e\Pi}}
\newcommand{\stmf}{\sigma^{(2)}_{\mu\Pi}}
\newcommand{\stxf}{\sigma^{(2)}_{x\Pi}\,\Bigl\{^+_-}
\newcommand{\stpf}{\sigma^{(2)}_{p\Pi}}

\begin{document}
\thispagestyle{empty}
\begin{flushright}
FR-PHENO-2023-08\\
IPPP/23/39\\
PSI-PR-23-27\\
ZU-TH 39/23
\end{flushright}
\vspace{2em}
\begin{center}
{\Large\bf Impact of NNLO QED corrections on \\[5pt] lepton-proton scattering at MUSE}
\\
\vspace{3em}
{\sc T.~Engel$^{\,a}$, F.~Hagelstein$^{\,b,\,c}$, M.~Rocco$^{\,c}$, V.~Sharkovska$^{\,c,\,d}$, A.~Signer$^{\,c,\,d}$, Y.~Ulrich$^{\,e}$
}\\[2em]
\vspace{0.3cm}
${}^a$ Albert-Ludwigs-Universität Freiburg, Physikalisches Institut, \\
Hermann-Herder-Straße 3, D-79104 Freiburg, Germany \\
\vspace{0.3cm}
${}^b$ Institute of Nuclear Physics \& PRISMA$^+$ Cluster of Excellence, \\Johannes Gutenberg-Universit\"at, 55099 Mainz, Germany\\
\vspace{0.3cm}
${}^c$ Paul Scherrer Institut,
CH-5232 Villigen PSI, Switzerland \\
\vspace{0.3cm}
${}^d$ Physik-Institut, Universit\"at Z\"urich,
CH-8057 Z\"urich, Switzerland \\
\vspace{0.3cm}
${}^e$ Institute for Particle Physics Phenomenology, Department of
Physics, \\
Durham University, Durham, DH1 3LE, UK\\

\setcounter{footnote}{0}
\end{center}
\vspace{2em}
\begin{abstract}
{} We present the complete next-to-next-to-leading order (NNLO) pure pointlike QED corrections to lepton-proton scattering, including three-photon-exchange contributions, and
investigate their impact in the case of the MUSE
experiment. These corrections are computed with no approximation regarding the
energy of the emitted photons and taking into account
lepton-mass effects. We contrast the NNLO QED corrections to known next-to-leading order corrections, where we include the elastic two-photon exchange (TPE) through a simple hadronic model calculation with a dipole ansatz for the proton electromagnetic form factors. We show that, in the low-momentum-transfer region accessed by the MUSE experiment, the improvement due to more
sophisticated treatments of the TPE, including inelastic TPE, is of similar if not smaller size than some of the NNLO QED corrections. Hence, the latter have to be included in a
precision determination of the low-energy proton structure from scattering
data, in particular for electron-proton scattering. For
muon-proton scattering, the NNLO QED corrections are considerably smaller.
\end{abstract}

\newpage
\setcounter{page}{1}

\section{Introduction} \label{sec:intro}

The scattering of electrons and muons off protons has been used for decades to obtain information on the structure of the proton. Still, in the regime of low 
energies, where the quark content of the proton is not yet resolved and the 
scattering is described with the help of form factors, there are several open 
questions and  discrepancies, see~\cite{Afanasev:2023gev} for a recent review. In view of this unsatisfactory situation it is 
important to revisit the theoretical aspects related to the extraction of form 
factors of the proton, with careful consideration of all effects that influence
the differential distributions of the final-state particles.
In addition to the uncertainty budget of radiative corrections due to hadronic contributions, dominated by the two-photon-exchange (TPE), this also includes standard QED corrections. 
The latter can lead to additional real photons in the final state, and a precise confrontation of theory with experiment needs to specify how such radiative events are treated. 

The analyses carried out so far have taken into account QED corrections at 
next-to-leading order (NLO), often with additional approximations~\cite{Mo:1968cg, Maximon:2000hm, Bystritskiy:2007hw, Kuraev:2013dra, Gramolin:2014pva, Gerasimov:2015aoa,Akushevich:2015toa}. However, perturbative calculations of 
QED corrections with pointlike particles to fully differential cross sections have now reached a maturity 
that allows to obtain complete next-to-next-to-leading order (NNLO) corrections to $2\to{2}$
processes~\cite{Bucoveanu:2018soy, Banerjee:2020rww,
  CarloniCalame:2020yoz, Banerjee:2021mty, Banerjee:2021qvi,
  Broggio:2022htr, Kollatzsch:2022bqa}. In the following, we refer to these types of corrections as \emph{pure} QED corrections. These computations can be done including mass effects and without making any 
approximation on the energy range of the emitted photons. This provides an 
opportunity to obtain unprecedented accuracy for the pure pointlike QED part of the
low-energy lepton-scattering processes. 

The presence of non-pointlike hadrons poses an additional challenge. While the emission of a single photon from an 
on-shell proton line can be described by two electromagnetic form factors, Feynman diagrams
with more complicated topologies, e.g.\ involving hadronic intermediate states, are more difficult to describe. Experimental analyses used to include the TPE as evaluated in the article by Mo-Tsai \cite{Mo:1968cg} or the later article by Maximon-Tjon \cite{Maximon:2000hm}. That is the elastic TPE, which has a proton in the intermediate state, as well as real radiation (bremsstrahlung), both in the limit of soft photons. The precision of modern scattering experiments --- take for example the A1 \cite{A1:2013fsc} and initial-state-radiation (ISR) \cite{Mihovilovic:2016rkr,Mihovilovic:2019jiz} experiments at MAMI or the PRad \cite{Xiong:2019umf} experiment at JLab --- required to go beyond that approximation and consider a more complete treatment of TPE and real radiation \cite{Vanderhaeghen:2000ws,Gramolin:2014pva}. Corrections beyond the soft-photon approximation are sometimes referred to as ``hard TPE'' and hard-photon radiation, respectively \cite{Afanasev:2023gev}. In the following, we refer to corrections from  diagrams with exchange of two virtual photons, shown in (\ref{eq:Mntpe}), as {\it virtual} TPE, and to the interference of one-photon-exchange (OPE) diagrams with a single bremsstrahlung photon radiated from the lepton and proton line, respectively, shown in (\ref{eq:Mn+1tpe}), as {\it real} TPE.

At low energies, the TPE contributions cannot be computed 
in perturbative QCD directly. They need to be modeled or, preferably, evaluated without model dependence in an effective-field theory framework \cite{Choudhary:2023rsz,Dye:2016uep,Dye:2018rgg} or through the use of dispersion relations. The latter require further experimental input, see~\cite{Tomalak:2016vbf,Tomalak:2017shs,Ahmed:2020uso} for a selection of recent data-driven evaluations.  Since the first works suggested an insufficiently precise description of the hard TPE as the origin of the discrepancy between form factor extractions from unpolarised and polarisation-transfer measurements \cite{Guichon:2003qm,Blunden:2003sp,Kondratyuk:2005kk,Chen:2004tw}, a vast literature on how to obtain and improve virtual TPE contributions appeared, see~\cite{Carlson:2007sp,Arrington:2011dn,Afanasev:2017gsk,Borisyuk:2019gym} for reviews focusing solely on virtual TPE in lepton-proton scattering.

The main aim of this investigation is not to
improve the predictions for TPE  as such but, rather, to critically assess
the impact of TPE corrections available in the literature, relative to
other corrections to lepton-proton scattering. To this end, we
 combine a simplified implementation of the TPE corrections  with 
state-of-the-art NNLO QED corrections.

We focus our application on the high-precision muon-scattering experiment MUSE~\cite{MUSE:2017dod, Cline:2021ehf}, which uses a beam of electrons and muons of both charges ($e^+$ and $\mu^+$ as well as $e^-$ and $\mu^-$), with  three different beam momenta:\footnote{Very recently, in~\cite{Li:2023sxf}, the beam momentum $153$ MeV was changed to $161$ MeV. In Figure \ref{fig:TPEBoth}, we  consider $153$ MeV in order to compare to older theory predictions for MUSE kinematics \cite{Tomalak:2015hva}.} $p_\text{beam}=115, 153, 210$ MeV. Its aim is to  compare extractions of the proton charge radius from electron and muon scattering, respectively, obtained with the same experimental setup, and to experimentally determine TPE corrections making use of both beam polarities.
The MUSE kinematics is limited to the low momentum-transfer region ($0.08$ GeV$^2$ for $p_\text{beam}=210$\,MeV), where
the TPE corrections are dominated by the elastic TPE, while the inelastic TPE is smaller than the anticipated accuracy of the MUSE cross-section measurements \cite{Tomalak:2015hva}. Therefore, as a reasonable first approximation in the MUSE kinematics, we implement a simple model for the elastic TPE contribution and neglect the inelastic part.

All considered corrections are implemented in the \mcmule{} framework~\cite{Banerjee:2020rww}. This goes beyond the NLO radiative corrections from \cite{Gramolin:2014pva} applied in the recent MUSE analysis of instrumental uncertainties \cite{Li:2023sxf}. In particular, we adapt the recent NNLO computation for muon-electron scattering~\cite{Broggio:2022htr} to obtain the NNLO QED corrections for lepton-proton scattering for pointlike protons in a fully differential way. It is the first time that these corrections (including three-photon exchange contributions) are taken into account, and we assess their relevance relative to variations in the treatment of TPE corrections. 

In Section~\ref{sec:calc} we will give a detailed description of the contributions that are included in our calculation.  This allows us to present in Section~\ref{sec:res} results for MUSE with $p_\text{beam}=210$\,MeV and study the impact of NNLO QED corrections. Our conclusions and an outlook towards further work will be presented in Section~\ref{sec:concl}.

\section{Calculation}  \label{sec:calc}

In order to maximally exploit the technical progress in the
computation of QED corrections, we take as the starting point a
pointlike interaction $ie \,q_p\,\gamma^\mu$ of
the photon with the proton, and we will call this the pure QED
contribution. We introduce the charge of the proton $q_p=1$ in units of $e$ for
bookkeeping purposes. The non-pointlike structure of the proton will be taken into account by additional contributions, denoted by $ F\times\delta^\mu$. Again $F$ has been introduced for bookkeeping purposes. Thus, for the
photon-proton interaction we write
\begin{align}\label{eq:phpr}
\begin{aligned}
 & ie \left[ \gamma^\mu F_1(Q^2) + \frac{i}{2 M} \sigma^{\mu\nu}q_\nu F_2(Q^2)\right] =
    \parbox{2cm}{\includegraphics[width=2cm]{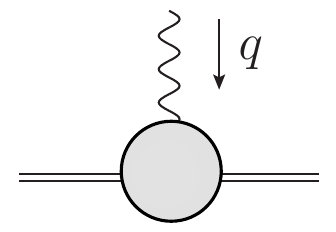}}
    \\
 = &\, ie \left[q_p\,\gamma^\mu + F\, \delta^\mu\right] =
 \parbox{2cm}{\includegraphics[width=2cm]{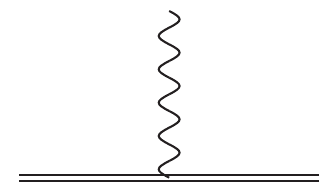}}
 +
 \parbox{2cm}{\includegraphics[width=2cm]{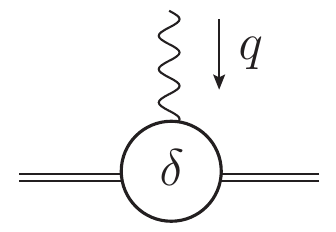}} 
\end{aligned}
\end{align}
with the four-momentum of the photon $q$, the spacelike virtuality of the photon $q^2=-Q^2<0$, the Dirac and Pauli form factors of the proton, $F_1(Q^2)$ and $F_2(Q^2)$, the proton mass $M$, and our notation for the antisymmetric combination of Dirac matrices $\sigma^{\mu \nu}=\frac{i}{2}\left(\gamma^\mu \gamma^\nu -\gamma^\nu \gamma^\mu\right)$. Note that for real photons, the form factors are normalised through their charge and anomalous magnetic moment $\kappa$: $F_1(0)=q_p=1$ and $F_2(0)=\kappa$.
In what follows we will describe in detail which contributions we
include, up to and including NNLO.

Starting at LO we obtain the matrix element (squared)
\begin{align}\label{eq:Mn0}
  \cM^{(0)}_n(q^2_\ell,F^2)&=\big|\cA^{(0)}_n(q_\ell,F)\big|^2 =
  \parbox{4cm}{\includegraphics[width=4cm]{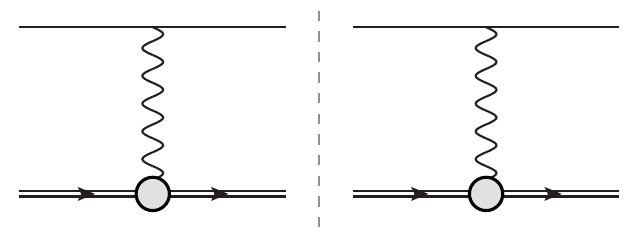}} 
\end{align}
by computing the tree-level amplitude of the $2\to{2}$ lepton-proton process with the full photon-proton vertex, depicted as
grey blobs in accordance with \eqref{eq:phpr}. The subscript $n$ indicates the number of final-state particles, i.e.~$n=2$ for the process considered in this paper. In the argument of the amplitude $\cA^{(0)}_n$ we indicate that
there is a single power of the coupling of the photon to the lepton, where
$q_\ell=\pm{1}$ is the charge of the positron or electron in units of $e$,
and that the full photon-proton vertex \eqref{eq:phpr} with arbitrary
form factors $F_1$ and $F_2$ is included. We suppress the dependence
on the external momenta but use the convention
\begin{align}\label{eq:extmom}
  \ell(p_1)\, p(p_2) \to \ell(p_3)\, p(p_4)\ + \{\gamma_i(k_i)\}
\end{align}
with up to two additional photons in the final state and either lepton polarity.

Thus, the unpolarised tree-level cross section is obtained by
integrating \eqref{eq:Mn0} over the two-body phase space $d\Phi_n$ and
including the standard flux and spin average factors
\begin{align}\label{eq:sigma0}
  d\sigma^{(0)}(q^2_\ell,F^2)&=
  \frac{1}{2s} \frac{1}{4} \int d\Phi_n  \cM^{(0)}_n(q^2_\ell,F^2)
  S(p_3,p_4) 
  \, .
\end{align}
The differential nature of the computation is coded in the measurement
function $S(p_3,p_4)$ that allows to include arbitrary cuts on the
final states. 

At NLO, virtual and real corrections contribute to the cross section. This leads to ultraviolet (UV) and infrared (IR) divergences. Both types of singularities are regularised dimensionally. For the UV divergences, we use the on-shell renormalisation scheme for
the masses and the coupling. The IR singularities cancel when
combining real and virtual corrections, as discussed in \cite{Afanasev:2023gev}. We perform the phase-space integrations numerically, using the FKS$^\ell$
subtraction method~\cite{Engel:2019nfw} to achieve this cancellation
for arbitrary IR-safe observables.

The so-called \textit{leptonic} corrections consist of Feynman diagrams with additional photons solely attached to the lepton line. This is a gauge-invariant subset of the complete NLO correction, and corresponds to the
OPE approximation. In the case of the electron, these corrections are expected to dominate due to collinear emission. This results in large logarithms of the form $\log(m_e^2/E^2)$ where the energy scale of the process $E$ is much larger than the electron mass $m_e$. Representative diagrams for the leptonic NLO corrections are
\begin{subequations}
\begin{align}\label{eq:Mn1l}
  \cM^{(1)}_n(q^4_\ell,F^2)&=
  2\,\Re\,\big(\cA^{(1)}_n(q^3_\ell,F)\, \cA^{(0)\,*}_n(q_\ell,F)\big)&
  &\supset&
  &\parbox{4cm}{\includegraphics[width=4cm]{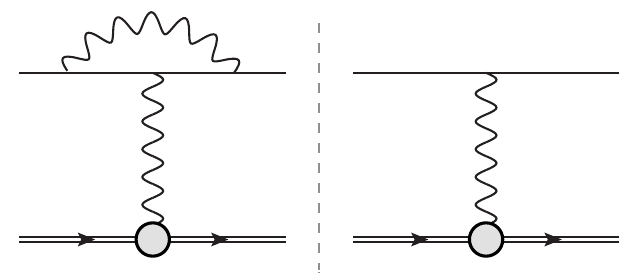}}& 
  \\ \label{eq:Mn+10l}
  \cM^{(0)}_{n+1}(q^4_\ell,F^2)&=
  \big|\cA^{(0)}_{n+1}(q^2_\ell,F)\big|^2&
  &\supset&
  &\parbox{4cm}{\includegraphics[width=4cm]{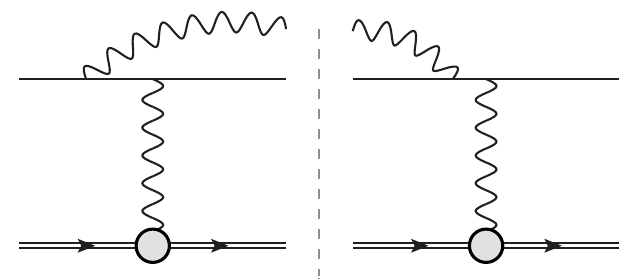}}&
\end{align}
\end{subequations}
The virtual corrections $d\sigma^{(1)}_v(q^4_\ell,F^2)$ are obtained
by integration of the one-loop matrix element
$\cM^{(1)}_n(q^4_\ell,F^2)$ 
over the two-body phase
space, whereas for the real corrections
$d\sigma^{(1)}_r(q^4_\ell,F^2)$ we have to integrate the matrix element $\cM^{(0)}_{n+1}(q^4_\ell,F^2)$ describing real radiation
over the three-body phase space
$d\Phi_{n+1}$. All these calculations can be performed with
arbitrary form factors~\cite{Mo:1968cg,Maximon:2000hm} using standard methods, and we obtain the leptonic NLO
corrections
\begin{align}\label{eq:NLOl}
  d\sigma^{(1)}(q^4_\ell,F^2)&=
    d\sigma^{(1)}_v(q^4_\ell,F^2) + d\sigma^{(1)}_r(q^4_\ell,F^2)
\end{align}
to the cross section.

In analogy to the leptonic corrections, the \textit{protonic} corrections
include emission solely from the proton line. Technically speaking,
they are also OPE contributions. Note that we do
not absorb the pure QED virtual corrections, i.e.~the vertex corrections, into the
form factors, since they are IR divergent. In general, we consider all QED reducible contributions as independent from the form factors. As we will see, in practice all this has
very limited impact, since these corrections are very small compared to the leptonic OPE, even for
$\ell=\mu$. This is due to the lack of logarithmic enhancements since collinear radiation only results in logarithms of the form $\log(M^2/E^2)$ with $M^2 \approx E^2$ for the energy scale considered here. Thus, in a standard OPE approach, these corrections are
often neglected. In the following, we take the protonic corrections into account in the pointlike proton approximation. Again, we have virtual and
real corrections
\begin{subequations}
\begin{align}\label{eq:Mn1p}
  \cM^{(1)}_n(q^2_\ell,q_p^4)&=
  2\,\Re\,\big(\cA^{(1)}_n(q_\ell,q_p^3)\, \cA^{(0)\,*}_n(q_\ell,q_p)\big)&
  &\supset&
  &\parbox{4cm}{\includegraphics[width=4cm]{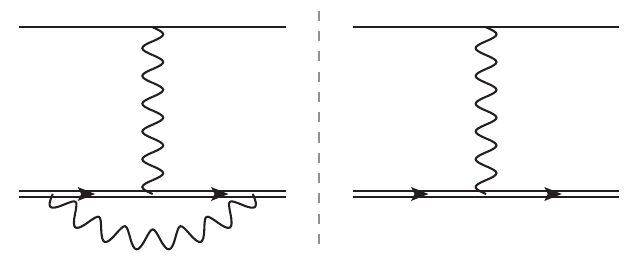}}& 
  \\ \label{eq:Mn+10p}
  \cM^{(0)}_{n+1}(q^2_\ell,q_p^4)&= 
  \big|\cA^{(0)}_{n+1}(q_\ell,q_p^2)\big|^2&
  &\supset&
  &\parbox{4cm}{\includegraphics[width=4cm]{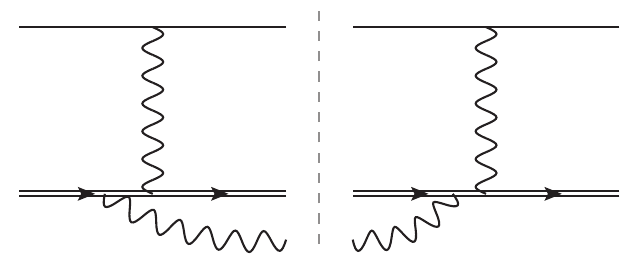}}&
\end{align}
\end{subequations}
where we only show a single diagram for illustrative purposes. 

The NLO corrections that involve additional photon couplings to both, the lepton and proton line, we call \textit{mixed} corrections or simply TPE. 
As a first approximation to the TPE at NLO, we consider the
elastic contribution which is due to an intermediate proton. For the virtual TPE
correction this results in box (and crossed box) diagrams whereas for
the real TPE corrections the intermediate proton is between the exchange
photon and the real  photon. Concretely, we take into account
\begin{subequations}
\label{eq:rvtpe}
\begin{align}\label{eq:Mntpe}
  \cM^{(1)}_n(q^3_\ell,F^3)&=
  2\,\Re\,\big(\cA^{(1)}_n(q^2_\ell,F^2)\, \cA^{(0)\,*}_n(q_\ell,F)\big)&
  &\supset&
  &\parbox{4cm}{\includegraphics[width=4cm]{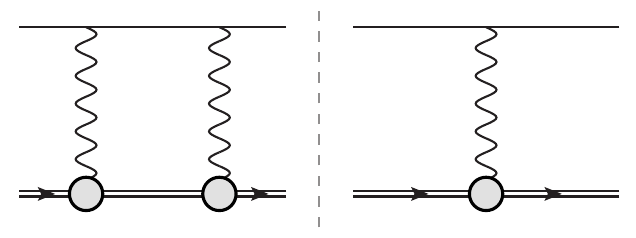}}& 
  \\ \label{eq:Mn+1tpe}
  \cM^{(0)}_{n+1}(q^3_\ell,F^3)&=
  2\,\Re\,\big(\cA^{(0)}_{n+1}(q_\ell,F^2)
  \cA^{(0)\,*}_{n+1}(q^2_\ell,F)\big)&
  &\supset& 
  &\parbox{4cm}{\includegraphics[width=4cm]{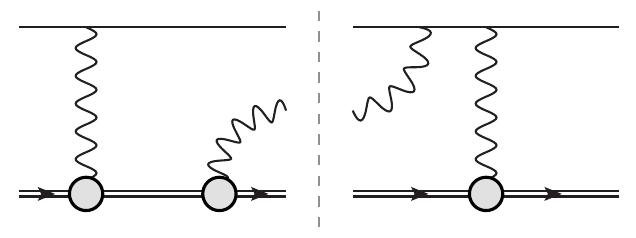}}&
\end{align}
\end{subequations}
We include the real TPE contribution
$d\sigma_r^{(1)}(q^3_\ell,F^3)$ by integrating
$\cM^{(0)}_{n+1}(q^3_\ell,F^3)$ 
over the three-body phase space $d\Phi_{n+1}$. Combining this with the corresponding 
virtual TPE correction $d\sigma_v^{(1)}(q^3_\ell,F^3)$ leads to an IR finite
result for any IR safe observable. 

\begin{figure}[t]
\begin{center}
\includegraphics[scale=0.55,angle=0]{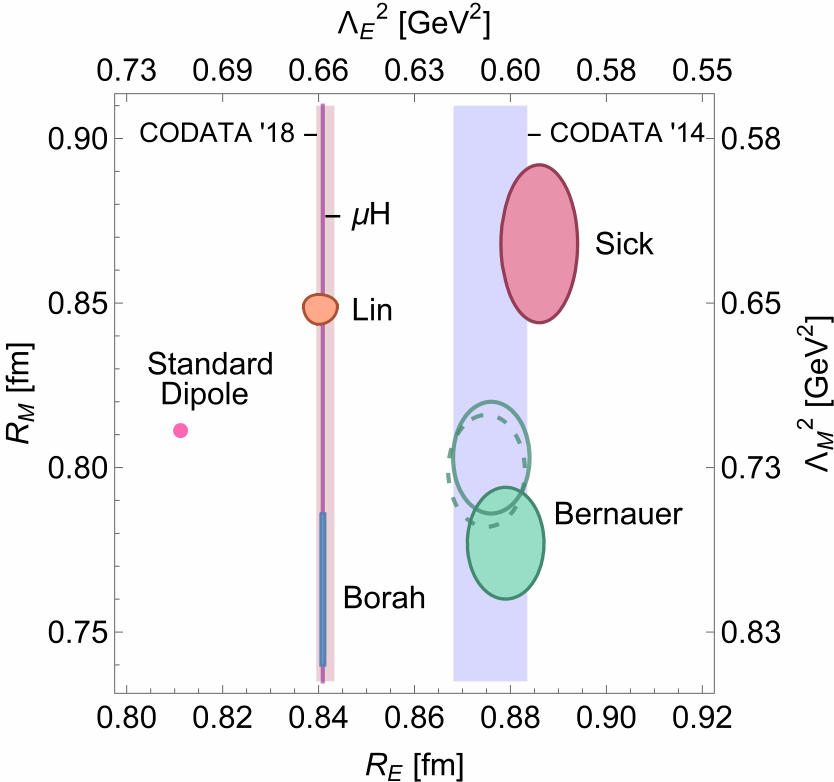}
\end{center}
\caption{
\label{fig:RERM}
Comparison of extractions of the proton electric and magnetic radii, $R_E$ and $R_M$, from different parametrisations of the proton Sachs form factors \cite{Borah:2020gte,Sick:2012zz,Lin:2021xrc,A1:2013fsc}, including the standard dipole with $\Lambda^2=0.71$ GeV$^2$, to the extraction from the muonic-hydrogen Lamb shift \cite{Antognini:2013txn}, and the CODATA recommended values for $R_E$, before (CODATA '14 \cite{Mohr:2015ccw}) and after (CODATA '18 \cite{Tiesinga:2021myr}) inclusion of the data from muonic hydrogen. Note that the displayed Bernauer results \cite{A1:2013fsc} are including hard TPE corrections from~\cite{Borisyuk:2006uq} (solid) and~\cite{Arrington:2011dn,Blunden:2005ew} (dashed), respectively. On the additional axes, we show which value of  $\Lambda_{E,M}^2$ would reproduce the radii if a dipole ansatz was assumed for the form factors.}
\end{figure}

\begin{figure}[ht!]
\begin{center}
\includegraphics[scale=0.75,angle=0]{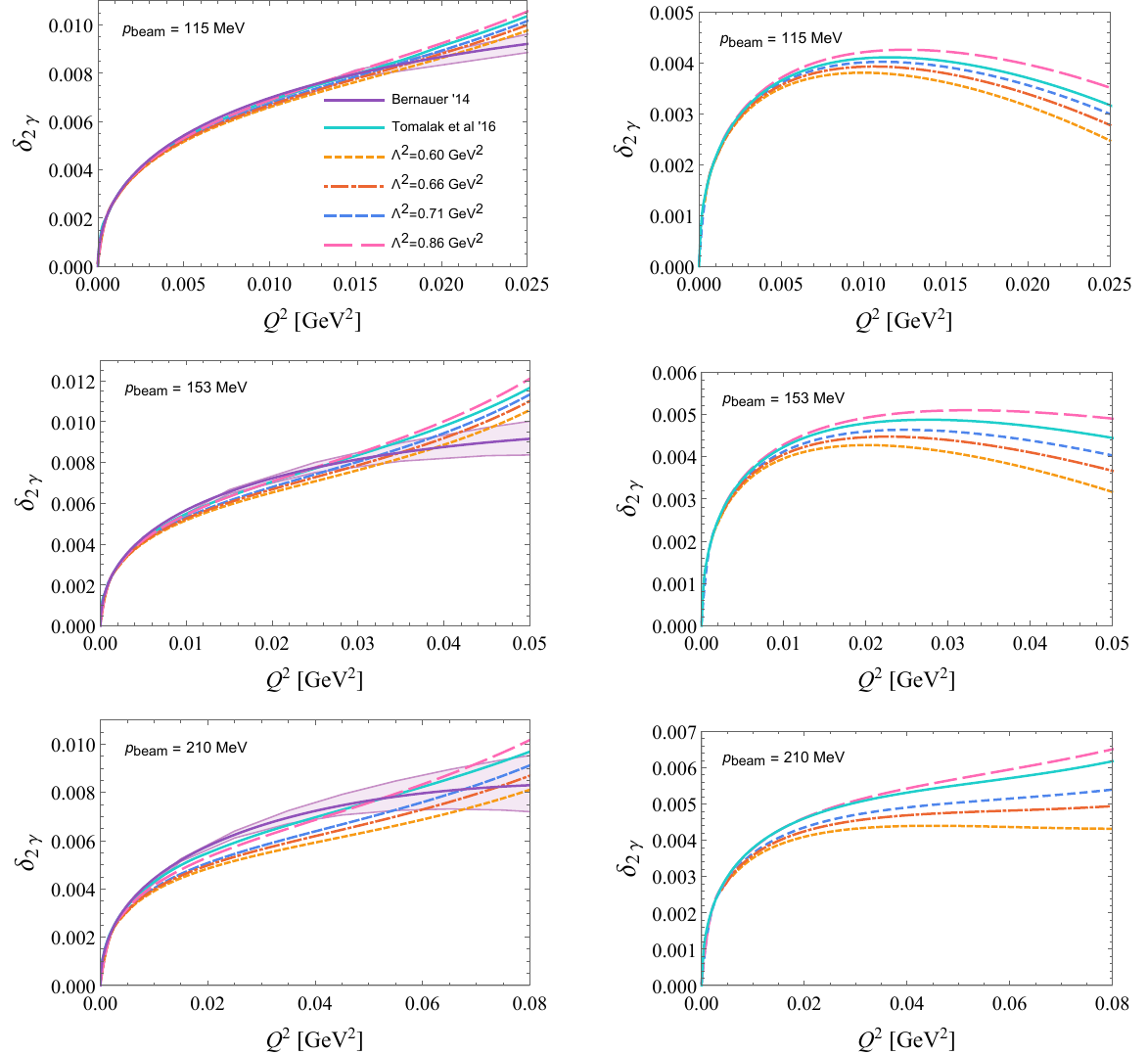}
\end{center}
\caption{
\label{fig:TPEBoth}
Comparison of virtual TPE corrections, $\delta_{2\gamma}$, to $e^- p$ (left column) and  $\mu^- p$ (right column) scattering for three different beam momenta envisaged by the MUSE experiment \cite{MUSE:2013uhu}: $p_\text{beam}=115, 153, 210$ MeV. The soft singularities are subtracted using the Maximon-Tjon prescription \cite{Maximon:2000hm}. Shown are the elastic TPE from our box model calculation with proton dipole form factors and different values of $\Lambda^2=0.60, 0.66, 0.71$ and $0.86$ GeV$^2$ (orange dotted, red dot-dashed, blue short-dashed and pink long-dashed lines), compared to the theoretical prediction for the total TPE from~\cite{Tomalak:2015hva} (solid cyan line), and the empirical extraction of the total TPE from~\cite{A1:2013fsc} (solid violet line with error band).}
\end{figure}

Besides the elastic TPE with a proton intermediate state, there is the so-called inelastic TPE contribution with inelastic intermediate states
\begin{align}\label{eq:MntpeX}
  &\mbox{virtual}
  \quad 
  \parbox{4cm}{\includegraphics[width=4cm]{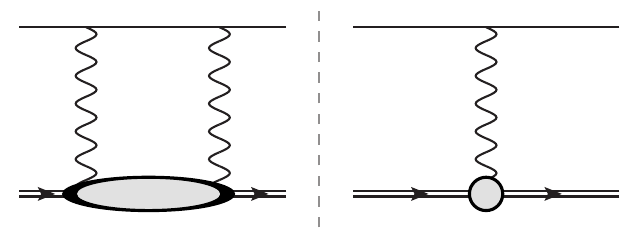}} &
  &\mbox{real}
  \quad 
  \parbox{4cm}{\includegraphics[width=4cm]{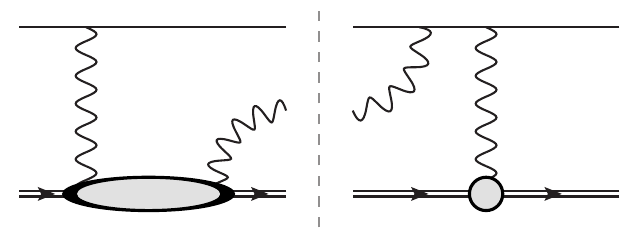}} &
\end{align}
The latter are denoted by the oval blobs in the above Feynman diagrams, and may contain anything: pions, resonances like the $\Delta(1232)$, etc. In the low-$Q^2$ region, relevant for the MUSE experiment, the elastic TPE dominates, while the inelastic TPE can be neglected. Therefore, the inelastic TPE is presently not included in \mcmule. Again, we leave this for a future update.

In this work, the virtual TPE has been implemented through a simple hadronic model calculation of the box and crossed-box diagrams in~\eqref{eq:Mntpe}, assuming on-shell proton form factors. The same approach has been used for instance in~\cite{Blunden:2003sp} and \cite{Tomalak:2014dja} for electron and muon scattering, respectively. In~\cite{Tomalak:2018jak}, the hadronic model calculation has been compared to a dispersive evaluation with one subtraction function. The latter involves an $s$-channel cut through the box diagram, thus, only needs input from on-shell form factors and does not require off-shell form factors, as the box calculation would in principle. Both approaches agree with $5\times 10^{-4}$ relative accuracy for the kinematics of muon scattering in the MUSE experiment \cite{Tomalak:2018jak}, and the same quality of the approximation can be assumed for electron scattering at MUSE. Thus, this model dependence in our approach can be safely neglected. For the proton electric and magnetic Sachs form factors
\begin{subequations} \label{eq:FF}
\begin{eqnarray}
    G_E(Q^2)&=&F_1(Q^2)-\frac{Q^2}{4\,M^2} F_2(Q^2),\\
    G_M(Q^2)&=&F_1(Q^2)+ F_2(Q^2),
\end{eqnarray}
we use a dipole ansatz
\begin{equation}
\label{eq:lambda}
G_E(Q^2)=G_D(Q^2)=\frac{G_M(Q^2)}{1+\kappa}, \qquad \text{with } \quad G_D(Q^2)=\left(\frac{\Lambda^2}{\Lambda^2+Q^2}\right)^2.
\end{equation}
\end{subequations}
Note that in the limit $\Lambda \to \infty$ the pure (pointlike) QED vertex is recovered. Of course, describing both the normalised electric and magnetic Sachs form factor through one single parameter $\Lambda$ is a rather naive ansatz. 
 Furthermore, the simple dipole form can only ever be a rough approximation to any form factor. Nevertheless, the standard dipole with $\Lambda^2=0.71$ GeV$^2$ is widely used as a reasonable first approximation to the proton form factors, and serves well our purpose to examine the relative importance of TPE and NNLO QED corrections. 
While we leave the implementation of the elastic TPE correction with input from modern form-factor parametrisations to a future version of \mcmule, we want to illustrate the impact of the form factors and the uncertainties in their description by considering various values for $\Lambda$. To motivate our choice, we consider the slopes of the Sachs form factors at $Q^2=0$, which are related to the charge and magnetisation radii of the proton 
\begin{equation}
    R_{E,M}=\sqrt{-\frac{6}{G_{E,M}(0)}\frac{dG_{E,M}(Q^2)}{dQ^2}}\Bigg\vert_{Q^2=0}
\end{equation}
shown in Figure~\ref{fig:RERM}.
Besides the standard dipole, we use  $\Lambda^2=0.86$ GeV$^2$ reproducing the small $R_M$ from~\cite{Borah:2020gte}, $\Lambda^2=0.66$ GeV$^2$ reproducing $R_E$ as extracted with unprecedented precision from the Lamb shift in muonic hydrogen \cite{Antognini:2013txn}, and $\Lambda^2=0.60$ GeV$^2$ reproducing the large $R_E$ from~\cite{Sick:2012zz}. 
The resulting impact on the virtual TPE correction, defined usually as
\begin{equation}
\label{eq:delta2gTPE}
\delta_{2\gamma}(\text{IR})   =   
\frac{\cM^{(1)}_{n}(q^3_\ell,F^3)\big|_\text{IR}}
{\cM^{(0)}_n(q^2_\ell,F^2)},
\end{equation}
is shown in Figure~\ref{fig:TPEBoth} for electron and muon scattering.
The ``IR" label indicates that the omission of $\cM^{(0)}_{n+1}(q^3_\ell,F^3)$ in (\ref{eq:delta2gTPE}) requires an unphysical subtraction of the IR singularity in $\cM^{(1)}_{n}(q^3_\ell,F^3)$. For the purpose of Figure~\ref{fig:TPEBoth} we have used the Maximon-Tjon subtraction~\cite{Maximon:2000hm, Afanasev:2023gev}.

  Going towards larger values of $Q^2$, excited intermediate states, e.g., resonances \cite{Ahmed:2020uso}, eventually do lead to sizeable contributions. In~\cite{Tomalak:2015hva}, the inelastic TPE correction has been estimated through a dispersive approach for near-forward kinematics, relating it to forward doubly-virtual Compton scattering amplitudes, which are in turn reconstructed dispersively with empirical input for the unpolarised proton structure functions. In their estimate, the inelastic TPE contribution, $\delta_{2\gamma}\sim 5\times 10^{-4}$, is smaller than the anticipated $1 \,\%$ accuracy of the cross-section measurements at MUSE.
Coincidentally, for $ep$ scattering, our evaluation of the elastic TPE  with a dipole parameter of $\Lambda^2=0.86$ GeV$^2$ is very close to their prediction for the total TPE~\cite{Tomalak:2015hva}.
This can be seen from Figure~\ref{fig:TPEBoth}, where we compare our spread of results  approximating the elastic TPE to the dispersive evaluation (solid cyan line) \cite{Tomalak:2015hva} and an empirical extraction (solid violet line and error band) \cite{A1:2013fsc} of the total TPE. Note that the elastic TPE included in  \cite{Tomalak:2015hva} agrees with our result for the standard dipole  (short-dashed blue line). The uncertainty on their inelastic TPE is small on the scale of the total TPE, and thus, omitted from the figure.

All contributions discussed so far were NLO. Moving towards NNLO, we
start again with the leptonic or OPE corrections. This gauge invariant
subset of NNLO corrections has been computed~\cite{Bucoveanu:2018soy,
  Banerjee:2020rww} for any choice of form factors. It consists of
double-virtual, real-virtual, and double-real corrections
\begin{align}\label{eq::sigma2}
  d\sigma^{(2)}(q^6_\ell,F^2)&=
    d\sigma^{(2)}_{vv}(q^6_\ell,F^2) + d\sigma^{(2)}_{rv}(q^6_\ell,F^2)
    + d\sigma^{(2)}_{rr}(q^6_\ell,F^2)
\end{align}
which are obtained by integrating the two-loop matrix element
$\cM^{(2)}_n(q^6_\ell,F^2)$ 
over $d\Phi_n$, the one-loop matrix element 
$\cM^{(1)}_{n+1}(q^6_\ell,F^2)$ 
over $d\Phi_{n+1}$, and the tree-level matrix element 
$\cM^{(0)}_{n+2}(q^6_\ell,F^2)$ 
over $d\Phi_{n+2}$, respectively. For any IR-safe observable, the IR
singularities of the individual parts in \eqref{eq::sigma2} cancel in
the sum. Representative Feynman diagrams of the various matrix
elements are
\begin{subequations}
\begin{align} \nonumber
  \cM^{(2)}_n(q^6_\ell,F^2)&= 
  2\,\Re\,\big(\cA^{(2)}_n(q^5_\ell,F)\, \cA^{(0)\,*}_n(q_\ell,F)\big)&
  &\supset&
  &\parbox{4cm}{\includegraphics[width=4cm]{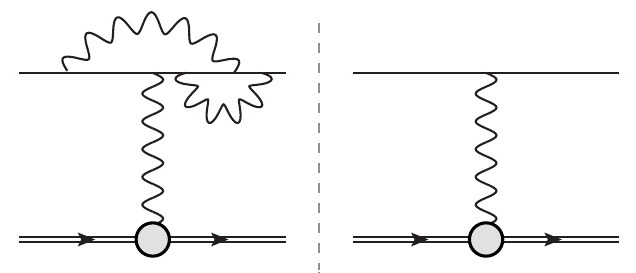}}& 
  \\\label{eq:Mn2l}
   & + \big|\cA^{(1)}_n(q^3_\ell,F)\big|^2&
  &\supset&
  &\parbox{4cm}{\includegraphics[width=4cm]{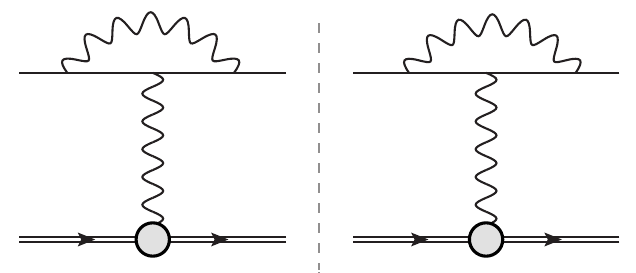}} &
  \\ \label{eq:Mn+11l}
  \cM^{(1)}_{n+1}(q^6_\ell,F^2)&=
  2\,\Re\,\big(\cA^{(1)}_{n+1}(q^4_\ell,F)
  \cA^{(0)\,*}_{n+1}(q^2_\ell,F)\big)&
  &\supset&
  &\parbox{4cm}{\includegraphics[width=4cm]{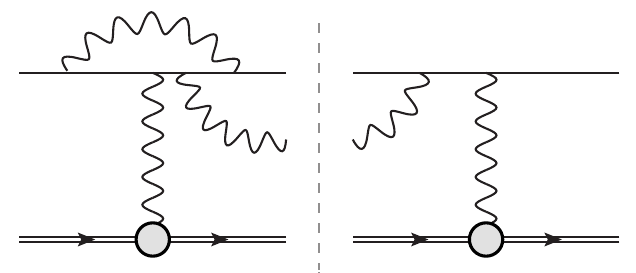}}&
  \\ \label{eq:Mn+20l}
  \cM^{(0)}_{n+2}(q^6_\ell,F^2)&=  
  \big|\cA^{(0)}_{n+2}(q^3_\ell,F)\big|^2&
  &\supset&
  &\parbox{4cm}{\includegraphics[width=4cm]{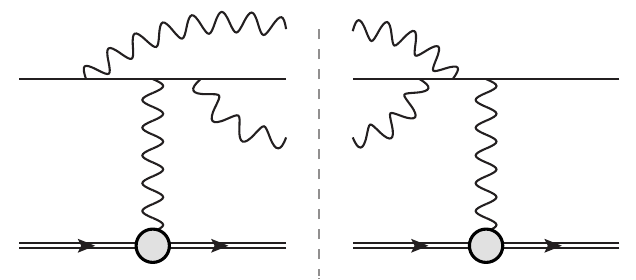}}\, . &
\end{align}
\end{subequations}
The one-loop amplitude squared, cf.\ the second line in \eqref{eq:Mn2l}, is included in the two-loop matrix element.
We note that some NLO diagrams for the process of $\ell p \to \ell p \gamma $, corresponding to an IR-finite subset of the $\ell p \to \ell p $ process at NNLO, have been previously included in \cite{Vanderhaeghen:2000ws} in approximate ways. The full set of leptonic NNLO corrections depicted in~\eqref{eq:Mn2l}, \eqref{eq:Mn+11l}, and \eqref{eq:Mn+20l} has been computed in~\cite{Bucoveanu:2018soy} with a slicing approach and later with the \mcmule{} framework in~\cite{Banerjee:2020rww}. The two results disagree substantially and a corresponding discussion can be found in~\cite{Banerjee:2020rww}.

The leptonic corrections are expected to be dominant, at least for the
case $\ell=e$, since they contain hard collinear emission from the
electron line. This leads to large logarithms. As we will see, the
size of these corrections depends crucially on the precise definition
of the observable. More concretely, the way additional photon radiation
is treated in the experiment will have a decisive impact. Thus, these
corrections have to be under control for empirical extractions of form factors and TPE effects.

The leptonic corrections are technically the most simple NNLO
corrections. Going beyond OPE, we have to consider one-loop pentagon
diagrams (for the real-virtual corrections) and, more challenging, a set
of topologically non-trivial two-loop diagrams, including (crossed)
double-box diagrams. In the language of lepton-proton scattering, they
correspond to three-photon exchange contributions. With current
techniques, it is not possible to do such a computation including form
factors. Hence, for all NNLO corrections beyond OPE we use the
approximation of a pointlike proton $\Lambda\to\infty$. In this case, the NNLO
corrections to lepton-proton scattering can be obtained from those of
muon-electron scattering, with adapted masses. The latter have been
computed~\cite{Broggio:2022htr,Engel:2022kde} using the two-loop
integrals of~\cite{Bonciani:2021okt}, as well as
OpenLoops~\cite{Buccioni:2019sur} and Package-X \cite{Patel:2015tea} for the one-loop amplitudes. They
can be split into gauge-invariant parts consisting of terms with a
fixed power of $q_\ell^n$ and $q_p^m$ with $n+m=8$. As an example, we illustrate the
virtual contributions to $d\sigma^{(2)}(q_\ell^4,q_p^4)$ which
requires the two-loop matrix element
\begin{subequations}
\begin{align} \nonumber
  \cM^{(2)}_n(q^4_\ell,q_p^4)=&
  2\,\Re\,\big(\cA^{(2)}_n(q^3_\ell,q_p^3)\, \cA^{(0)\,*}_n(q_\ell,q_p)\big)&
  &\supset&
  &\parbox{4cm}{\includegraphics[width=4cm]{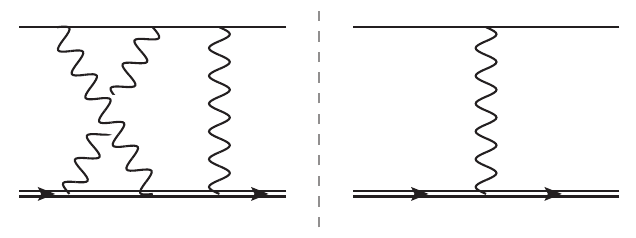}}& 
  \\ \label{eq:Mn2x}
   & + \big|\cA^{(1)}_n(q^2_\ell,q_p^2)\big|^2&
  &\supset&
  &\parbox{4cm}{\includegraphics[width=4cm]{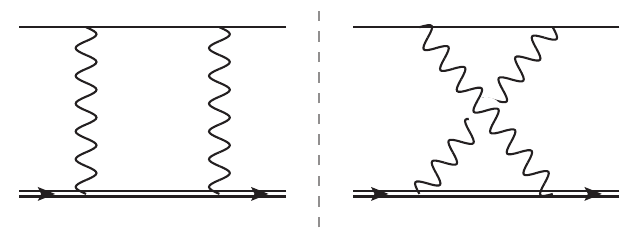}} &
  \\ \nonumber
  & + 2\,\Re\,\big(\cA^{(1)}_{n}(q^3_\ell,q_p)
  \cA^{(1)\,*}_{n}(q_\ell,q_p^3)\big)&
  &\supset&
  &\parbox{4cm}{\includegraphics[width=4cm]{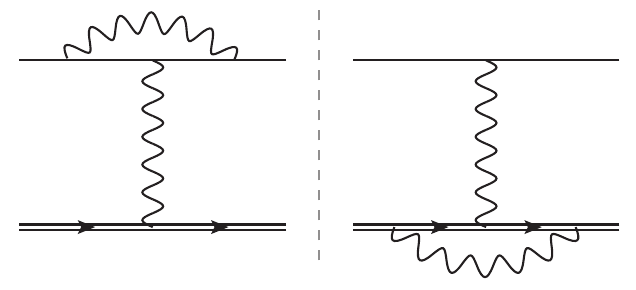}}&
\end{align}
\end{subequations}
Of course, real-virtual and double-real corrections have to be
considered as well. Representative examples for $d\sigma^{(2)}_{rv}(q_\ell^5,q_p^3)$ and $d\sigma^{(2)}_{rr}(q_\ell^5,q_p^3)$ are
\begin{subequations}
\begin{align}\label{eq:MnXrv}
  \cM^{(1)}_{n+1}(q^5_\ell,q_p^3)&=
  2\,\Re\,\big(\cA^{(1)}_{n+1}(q_\ell^3,q_p^2)\, \cA^{(0)\,*}_{n+1}(q_\ell^2,q_p)\big)&
  &\supset&
  &\parbox{4cm}{\includegraphics[width=4cm]{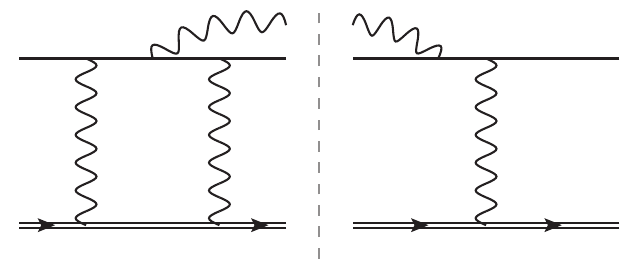}}& 
  \\ \label{eq:MnXrr}
  \cM^{(0)}_{n+2}(q^5_\ell,q_p^3)&= 
  2\,\Re\,\big(\cA^{(0)}_{n+2}(q^3_\ell,q_p)\, \cA^{(0)\,*}_{n+1}(q^2_\ell,q^2_p) \big)&
  &\supset&
  &\parbox{4cm}{\includegraphics[width=4cm]{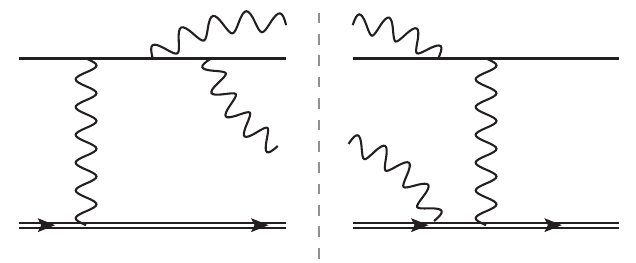}}&
\end{align}
\end{subequations}

All contributions considered so far are collectively called \textit{photonic} corrections. In addition, there are vacuum polarisation contributions. We include
electron, muon, tau loops in the vacuum polarisation $\Pi$, and collectively refer to them as \textit{fermionic} corrections. Note that we also include hadronic contributions in $\Pi$, however, they are about a factor 100 smaller than the fermionic  contributions. Vacuum polarization starts to contribute at NLO through
virtual effects. At NNLO, there are virtual and real fermionic corrections to be
included. In analogy to the other corrections, we use a form factor
for the OPE contributions and a pointlike proton interaction for
TPE. Sample diagrams for the virtual corrections are
\begin{subequations}
\begin{align}\label{eq:MnPI}
  \cM^{(1)}_n(q^2_\ell,\Pi, F^2)&=
  2\,\Re\,\big(\cA^{(1)}_n(q_\ell,\Pi, F)\, \cA^{(0)\,*}_n(q_\ell,F)\big)&
  &\supset&
  &\parbox{4cm}{\includegraphics[width=4cm]{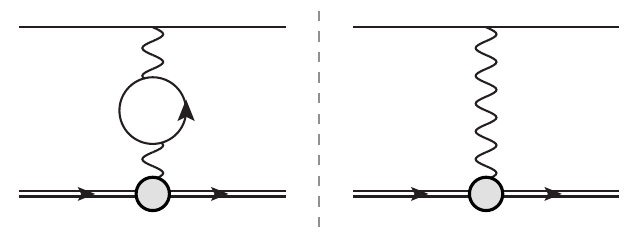}}& 
  \\ \label{eq:MnPIv}
  \cM^{(2)}_n(q^4_\ell,\Pi,F^2)&=
  2\,\Re\,\big(\cA^{(2)}_n(q^3_\ell,\Pi,F)\, \cA^{(0)\,*}_n(q_\ell,F)\big)&
  &\supset&
  &\parbox{4cm}{\includegraphics[width=4cm]{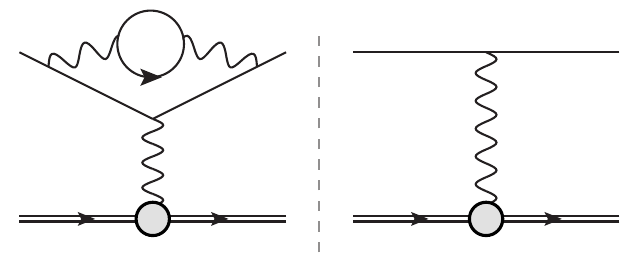}}& 
  \\ \label{eq:MnPIb}
  \cM^{(2)}_n(q^3_\ell,\Pi,q_p^3)&=
  2\,\Re\,\big(\cA^{(2)}_n(q^2_\ell,\Pi,q_p^2)\, \cA^{(0)\,*}_n(q_\ell,q_p)\big)&
  &\supset&
  &\parbox{4cm}{\includegraphics[width=4cm]{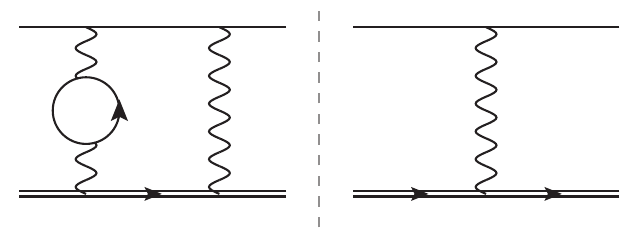}}& 
\end{align}
\end{subequations}
There are also contributions with either two one-loop insertions of $\Pi$ or a single insertion of a two-loop vacuum polarisation. They contribute to virtual corrections only and we denote them collectively by 
$d\sigma^{(2)}(q_\ell^2,\Pi^2,F^2)$. In the case of leptons, the analytic form of the two-loop vacuum polarisation from~\cite{Djouadi:1993ss} is used, whereas for hadronic loops, the Fortran library {\tt alphaQED}~\cite{alphaqed} is employed. We do not include real corrections with an additional $e^+\,e^-$ pair in the final state \cite{Heller:2021mcw}. This is a measurably different process. However, depending on the details of the experimental analysis, this process can contribute to lepton-proton cross sections at NNLO.

To summarise, our results for the NNLO cross section
\begin{align}\label{eg:xsnnlo}
  d\sigma_2&=d\sigma^{(0)}+d\sigma^{(1)}+d\sigma^{(2)} 
\end{align}
include 
\begin{subequations}
\label{eq:res}
\begin{align}
  d\sigma_0&=d\sigma^{(0)}(q_\ell^2,F^2), \\
  d\sigma^{(1)}&=d\sigma^{(1)}(q_\ell^4,F^2)
  +d\sigma^{(1)}(q_\ell^3,F^3) + d\sigma^{(1)}(q_\ell^2,q_p^4)   
  +d\sigma^{(1)}(q_\ell^2,\Pi,F^2), 
  \\
  d\sigma^{(2)}&=d\sigma^{(2)}(q_\ell^6,F^2) 
   + \Big(\sum_{j=3}^{5} d\sigma^{(2)}(q_\ell^j,q_p^{8-j}) \Big)
   + d\sigma^{(2)}(q_\ell^2,q_p^6)\nonumber \\
  & \phantom{=}
  +\Big(d\sigma^{(2)}(q_\ell^4,\Pi,F^2) + d\sigma^{(2)}(q_\ell^2,\Pi^2,F^2)\Big)
  +d\sigma^{(2)}(q_\ell^3,\Pi,q_p^3) 
  +d\sigma^{(2)}(q_\ell^2,\Pi,q_p^4),
\end{align}
\end{subequations}
where all parts are individually gauge independent. All contributions proportional to $F^2$ can easily be computed with arbitrary form factors. The term 
$d\sigma^{(1)}(q_\ell^3,F^3)$ has been computed using a dipole ansatz for the electromagnetic form factors. For the remaining terms we use the pointlike proton approximation. 

In Section~\ref{sec:res} we will present results for different values of the dipole parameter $\Lambda$ in \eqref{eq:lambda}, including $\Lambda=\infty$ for pointlike protons. In order to indicate the dependence on $\Lambda$, we will use the compact notation
\begin{subequations}
\label{eq:rescompact}
\begin{align}
  d\sigma_0^\Lambda&=d\sigma^{(0)}(q_\ell^2,F(\Lambda)^2), \\
  d\sigma^{(1)\Lambda}&=
  d\sigma_\ell^{(1)\Lambda}
  +d\sigma_{x}^{(1)\Lambda}
  +d\sigma_{p}^{(1)\infty}
  +d\sigma_{\Pi}^{(1)\Lambda},
  \\
  d\sigma^{(2)\Lambda}&=
  d\sigma_\ell^{(2)\Lambda}
  +d\sigma_x^{(2)\infty}
  +d\sigma_p^{(2)\infty}
  +d\sigma_{\ell\Pi}^{(2)\Lambda}
  +d\sigma_{x\Pi}^{(2)\infty}
  +d\sigma_{p\Pi}^{(2)\infty},
\end{align}
\end{subequations}
where the terms in \eqref{eq:rescompact} are in one-to-one correspondence with those of \eqref{eq:res}. The labels $\ell\in\{e,\mu\}$, $p$, and $x$ stand for \textit{leptonic} (i.e.~\textit{electronic} or \textit{muonic}), \textit{protonic}, and \textit{mixed} corrections. The terms $\sim q_\ell^2\Pi^2 F^2$ have been absorbed into $d\sigma_{\ell\Pi}^{(2)\Lambda}$ as a matter of convention. If the proton is treated pointlike, we set $\Lambda=\infty$ in the notation. Otherwise, we use the label $\Lambda\in\{60,71,86\}$ to indicate the value of the dipole parameter that has been used, where e.g. the label $\Lambda=60$ corresponds to $\Lambda^2=0.60\,\mathrm{GeV}^2$.

\section{Results and discussion} \label{sec:res}

This section presents results for lepton-proton scattering,
tailored to the characteristics of the MUSE experiment~\cite{MUSE:2017dod, Cline:2021ehf}.
Particular emphasis is given to the impact of NNLO pure QED corrections compared to hadronic effects at NLO, focusing mainly on the elastic TPE discussed in Section~\ref{sec:calc}.
The code employed for this study is publicly available at
\begin{center}
   \url{https://gitlab.com/mule-tools/mcmule}
\end{center}
and the whole set of results can be found in the relevant directory of
the \mcmule{} user library
\begin{center}
    \url{https://mule-tools.gitlab.io/user-library/}
\end{center}
along with user, menu and configuration files, and the Python code
that generates the plots in the paper~\cite{McMule:data}. The
production runs employed version {\tt v0.5.0} of the
\mcmule{} public release.

The kinematics of the process is defined by the momenta
in~(\ref{eq:extmom}) together with the lepton and the proton mass, $m_\ell$
and $M$.
Both polarities of the lepton are considered. 
For the purpose of illustration, we consider an incoming lepton of
momentum 
\begin{equation}
    p_{\rm beam}=|\vec{p}_1|=210\,\mathrm{MeV}
\end{equation}
scattering off a proton at rest. This
is consistent with one of the MUSE setups, and corresponds to a
centre-of-mass energy of $\sqrt{s}\approx 1.2$ GeV. The experimental
setup defines a window for the lepton scattering angle, $20^\circ <
\theta_\ell < 100^\circ$, and for the lepton final-state momentum
\begin{equation}
    |\vec{p}_3| > 15\,\mathrm{MeV} \equiv p_{\rm min}\, .
\end{equation}
All the results use the input parameters~\cite{Workman:2022ynf}
\begin{align*}
    \alpha &= 1/137.035999084      \,,&  m_e &=  0.510998950 \,\,{\rm MeV} \,, \\
    m_\mu  & =  105.658375 \,\,{\rm MeV} \,,&  M &= 938.2720813     \,\,{\rm MeV} \,.
\end{align*}
The most recent version {\tt alphaQEDc19} of the Fortran library {\tt alphaQED} is used for the evaluation
of diagrams with hadronic loop insertions.

In order to measure elastic scattering, kinematical cuts have to be applied to suppress radiative events. In the case of the MUSE experiment this is done by means of a forward-angle calorimeter~\cite{Li:2023sxf}. In the following, results are presented for two different scenarios,
{\tt S0} and {\tt S1}, depending on whether an additional inelasticity cut is enforced. In our
simulation, the energy of photon(s) emitted in the lab frame within a 100 mrad angle is
cumulated into $E_\gamma$. In scenario {\tt S1}, if $E_\gamma > 0.4\,p_{\rm beam}$ the
corresponding event is removed from the analysis. The kinematical details discussed above and used in our scenarios are summarised in
Table~\ref{tab:scenarios}.
\begin{table}[h]
    \centering
    \renewcommand{\arraystretch}{1.3}
    \begin{tabular}{l|c|c|c}
      & $20^\circ < \theta_\ell < 100^\circ$
      & $|\vec{p}_3| > p_{\rm min}$
      & $E_\gamma < 0.4\,p_{\rm beam}$  \\
        \hline
        {\tt S0}  & \checkmark & \checkmark  &  \\
        \hline
        {\tt S1}  & \checkmark & \checkmark  & \checkmark
    \end{tabular}
    \caption{Kinematical scenarios analysed in the \mcmule{}
      prediction.}
    \label{tab:scenarios}
\end{table}

\begin{figure}[h!]
  \centering
  \begin{minipage}{.49\textwidth}
    \centering
    \includegraphics[width=1.\textwidth]{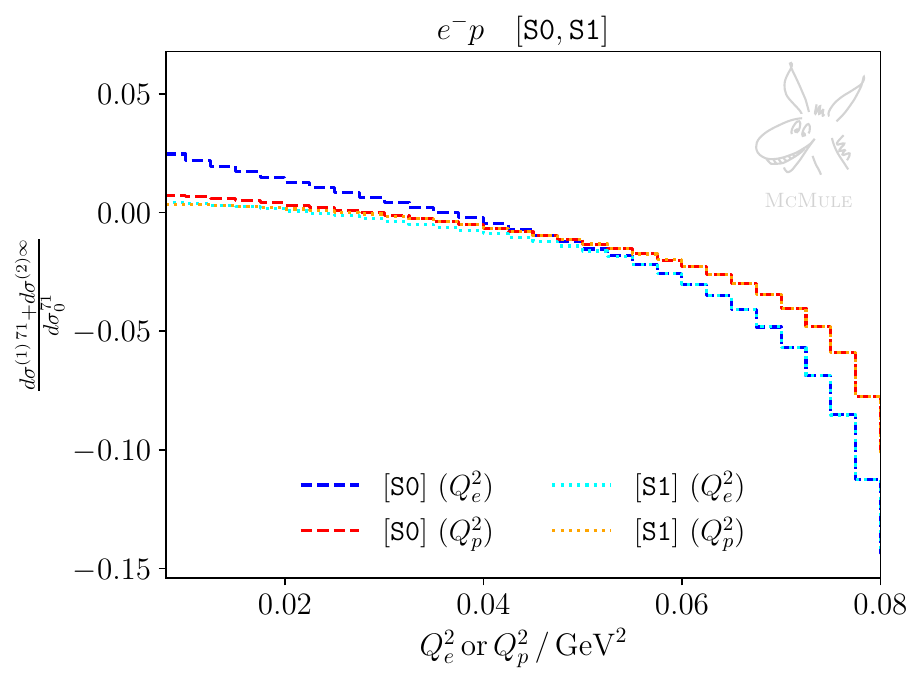}
  \end{minipage}
  \begin{minipage}{.49\textwidth}
    \centering
    \includegraphics[width=1.\textwidth]{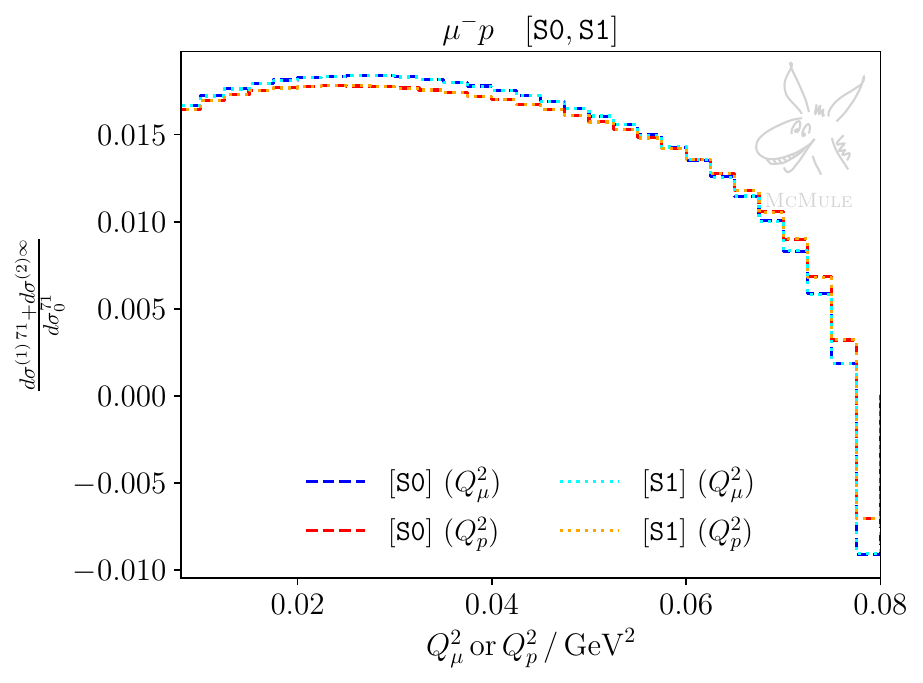}
  \end{minipage}
  \caption{Complete NLO+NNLO corrections for $ep$ (left panel)  and $\mu p$ (right panel) scattering, normalised to the Born cross sections. Shown are both kinematic scenarios
    {\tt S0} and {\tt S1}. Since
  corrections due to real-photon emissions are included, the distributions differ
  whether they are plotted as functions of $Q^2_\ell$ or $Q^2_p$. }
  \label{fig:qsql-diff}
\end{figure}

Figure~\ref{fig:qsql-diff} illustrates the impact of this forward-angle inelasticity cut on $E_\gamma$. It depicts the sum of the NLO and NNLO corrections normalised to the Born cross section as a function of the leptonic momentum transfer $Q_\ell^2 = -(p_1-p_3)^2$ as well as its protonic counterpart $Q_p^2=-(p_2-p_4)^2$. Results for both scenarios {\tt S0} and {\tt S1} are shown. In the presence of radiation we have $Q_\ell^2 \neq Q_p^2$ and the deviation of the two curves can be taken as a measure of inelastic effects. In the case of $\mu p$ scattering hard radiation is not collinearly enhanced due to the larger lepton mass. For $ep$ scattering, on the other hand, sizable deviations can be observed. We therefore restrict the discussion to this more interesting case.

For small momentum transfer, the {\tt S1} distributions w.r.t.\ $Q_e^2$ and $Q_p^2$ are almost identical. Hence, the $E_\gamma$ cut is able to remove most hard photon radiation in this region. This is not the case for larger momentum transfer where the $Q_e^2$ and $Q_p^2$ curves start to deviate. This behaviour can be understood as follows. Small momentum transfer corresponds to forward scattering of the lepton. In this case both initial-state as well as final-state collinear radiation is emitted in forward direction and therefore removed by the $E_\gamma$ cut. For larger $Q^2$, on the other hand, final-state collinear radiation is not forwardly directed and is thus not vetoed. Nevertheless, the inelasticity cut still removes initial-state collinear radiation. However, this effect seems to be very small since the {\tt S0} and {\tt S1} scenarios approach each other in this region. Thus, final-state radiation dominates the inelastic effects for larger momentum transfer irrespective of the cut on $E_\gamma$.

The order-by-order contributions, $\sigma^{(i)}$, to the LO, NLO, and NNLO integrated cross section are
shown in Tables~\ref{tab:tot_xsec_e} and~\ref{tab:tot_xsec_m} for $ep$
and $\mu p$ scattering, respectively. The various contributions are denoted as in \eqref{eq:rescompact}. The results are presented for
both kinematical scenarios and for different values of $\Lambda$
entering the dipole ansatz of the proton form factors~(\ref{eq:lambda}). In particular,
$\Lambda^2\in\{0.60,\,0.71,\,0.86\}$ GeV$^2$, while the label $\Lambda\infty$
stands for a pure pointlike QED photon-proton vertex. The NNLO corrections have been evaluated only for the point-like proton ($\Lambda\infty$) and the standard dipole form factors ($\Lambda71$). We reiterate that inelastic contributions to TPE are not taken into account in the calculation.

\begin{table}[h!]
    \centering
    \renewcommand{\arraystretch}{1.19}
    \resizebox{\textwidth}{!}{%
    \begin{tabular}{l|rrrr|rrrr}
        &\multicolumn{4}{ c}{$\sigma / \upmu{\rm b}\,\,[{\tt S0}]$}
        &\multicolumn{4}{|c}{$\sigma / \upmu{\rm b}\,\,[{\tt S1}]$} \\
        & $\Lambda\infty\phantom{a}$ & $\Lambda60\phantom{a}$
        & $\Lambda71    \phantom{a}$ & $\Lambda86\phantom{a}$
        & $\Lambda\infty\phantom{a}$ & $\Lambda60\phantom{a}$
        & $\Lambda71    \phantom{a}$ & $\Lambda86\phantom{a}$ \\
        \hline
        $\szet  $&$ 40.6564 $&$ 38.5302 $&$ 39.0482 $&$ 39.5432 $
                 &$ 40.6564 $&$ 38.5302 $&$ 39.0482 $&$ 39.5432 $\\
        \hline
        $\soel  $&$  6.3603 $&$  6.3721 $&$  6.3705 $&$  6.3687 $
                 &$  0.9438 $&$  0.9735 $&$  0.9672 $&$  0.9610 $\\
        \multirow{2}{*}{
        $\soxl$} &$ -0.1931 $&$ -0.1526 $&$ -0.1609 $&$ -0.1696 $
                 &$ -0.1924 $&$ -0.1520 $&$ -0.1603 $&$ -0.1689 $\\
                 &$  0.1931 $&$  0.1526 $&$  0.1609 $&$  0.1696 $
                 &$  0.1924 $&$  0.1520 $&$  0.1603 $&$  0.1689 $\\
        $\sopl  $&$ -0.0020 $&$         $&$         $&$         $
                 &$ -0.0020 $&$         $&$         $&$         $\\
        $\solf  $&$  0.5878 $&$  0.5554 $&$  0.5634 $&$  0.5711 $
                 &$  0.5878 $&$  0.5554 $&$  0.5634 $&$  0.5711 $\\
        \hline
        $\stel  $&$ -0.0134 $&$         $&$ -0.0080 $&$         $
                 &$ -0.0102 $&$         $&$ -0.0049 $&$         $\\
        \multirow{2}{*}{
        $\stxl$} &$ -0.0240 $&$         $&$         $&$         $
                 &$ -0.0009 $&$         $&$         $&$         $\\
                 &$  0.0279 $&$         $&$         $&$         $
                 &$  0.0049 $&$         $&$         $&$         $\\
        $\stpl  $&$ -0.0000 $&$         $&$         $&$         $
                 &$ -0.0000 $&$         $&$         $&$         $\\
        $\stef  $&$  0.0540 $&$         $&$  0.0542 $&$         $
                 &$  0.0094 $&$         $&$  0.0098 $&$         $\\
        \multirow{2}{*}{
        $\stxf$} &$ -0.0046 $&$         $&$         $&$         $
                 &$ -0.0046 $&$         $&$         $&$         $\\
                 &$  0.0046 $&$         $&$         $&$         $
                 &$  0.0046 $&$         $&$         $&$         $\\
        $\stpf  $&$ -0.0001 $&$         $&$         $&$         $
                 &$ -0.0001 $&$         $&$         $&$         $
    \end{tabular}
    }
    \caption{Integrated cross sections for $ep$ scattering, for both
      {\tt S0} and {\tt S1} scenarios, at LO, NLO, and NNLO. All digits shown are significant. $\Lambda\infty$ denotes pure QED contributions with a pointlike proton, $\{\Lambda60,\,\Lambda71,\,\Lambda86\}$ stand for proton finite-size corrections with proton form factors modeled through a dipole ansatz and $\Lambda^2$ set to $\{0.60,\,0.71,\,0.86\}$
      GeV$^2$, respectively. When applicable, the different
      contributions with positive and negative electrons are shown. }
    \label{tab:tot_xsec_e}
\end{table}

\begin{table}[h!]
    \centering
    \renewcommand{\arraystretch}{1.19}
    \resizebox{\textwidth}{!}{%
    \begin{tabular}{l|rrrr|rrrr}
        &\multicolumn{4}{ c}{$\sigma / \upmu{\rm b}\,\,[{\tt S0}]$}
        &\multicolumn{4}{|c}{$\sigma / \upmu{\rm b}\,\,[{\tt S1}]$} \\
        & $\Lambda\infty\phantom{a}$ & $\Lambda60\phantom{a}$
        & $\Lambda71    \phantom{a}$ & $\Lambda86\phantom{a}$
        & $\Lambda\infty\phantom{a}$ & $\Lambda60\phantom{a}$
        & $\Lambda71    \phantom{a}$ & $\Lambda86\phantom{a}$ \\
        \hline
        $\szet  $&$ 52.1775 $&$ 49.0046 $&$ 49.6678 $&$ 50.3161 $
                 &$ 52.1775 $&$ 49.0046 $&$ 49.6678 $&$ 50.3161 $\\
        \hline
        $\soml  $&$ -0.0710 $&$ -0.0613 $&$ -0.0631 $&$ -0.0649 $
                 &$ -0.0713 $&$ -0.0616 $&$ -0.0634 $&$ -0.0652 $\\
        \multirow{2}{*}{
        $\soxl$} &$ -0.2196 $&$ -0.1594 $&$ -0.1703 $&$ -0.1817 $
                 &$ -0.2196 $&$ -0.1594 $&$ -0.1703 $&$ -0.1817 $\\
                 &$  0.2196 $&$  0.1594 $&$  0.1703 $&$  0.1817 $
                 &$  0.2196 $&$  0.1594 $&$  0.1703 $&$  0.1817 $\\
        $\sopl  $&$ -0.0034 $&$         $&$         $&$         $
                 &$ -0.0034 $&$         $&$         $&$         $\\
        $\solf  $&$  0.7557 $&$  0.7070 $&$  0.7172 $&$  0.7273 $
                 &$  0.7557 $&$  0.7070 $&$  0.7172 $&$  0.7273 $\\
        \hline
        $\stml  $&$ -0.0000 $&$         $&$ -0.0001 $&$         $
                 &$ -0.0000 $&$         $&$ -0.0001 $&$         $\\
        \multirow{2}{*}{
        $\stxl$} &$  0.0010 $&$         $&$         $&$         $
                 &$  0.0010 $&$         $&$         $&$         $\\
                 &$  0.0037 $&$         $&$         $&$         $
                 &$  0.0037 $&$         $&$         $&$         $\\
        $\stpl  $&$ -0.0000 $&$         $&$         $&$         $
                 &$ -0.0000 $&$         $&$         $&$         $\\
        $\stmf  $&$  0.0079 $&$         $&$  0.0076 $&$         $
                 &$  0.0079 $&$         $&$  0.0076 $&$         $\\
        \multirow{2}{*}{
        $\stxf$} &$ -0.0057 $&$         $&$         $&$         $
                 &$ -0.0057 $&$         $&$         $&$         $\\
                 &$  0.0057 $&$         $&$         $&$         $
                 &$  0.0057 $&$         $&$         $&$         $\\
        $\stpf  $&$ -0.0001 $&$         $&$         $&$         $
                 &$ -0.0001 $&$         $&$         $&$         $
    \end{tabular}
    }
    \caption{Same as Table \ref{tab:tot_xsec_e} but for $\mu p$ scattering.}
    \label{tab:tot_xsec_m}
\end{table}

\begin{figure}
  \centering
  \includegraphics[width=1.\textwidth]{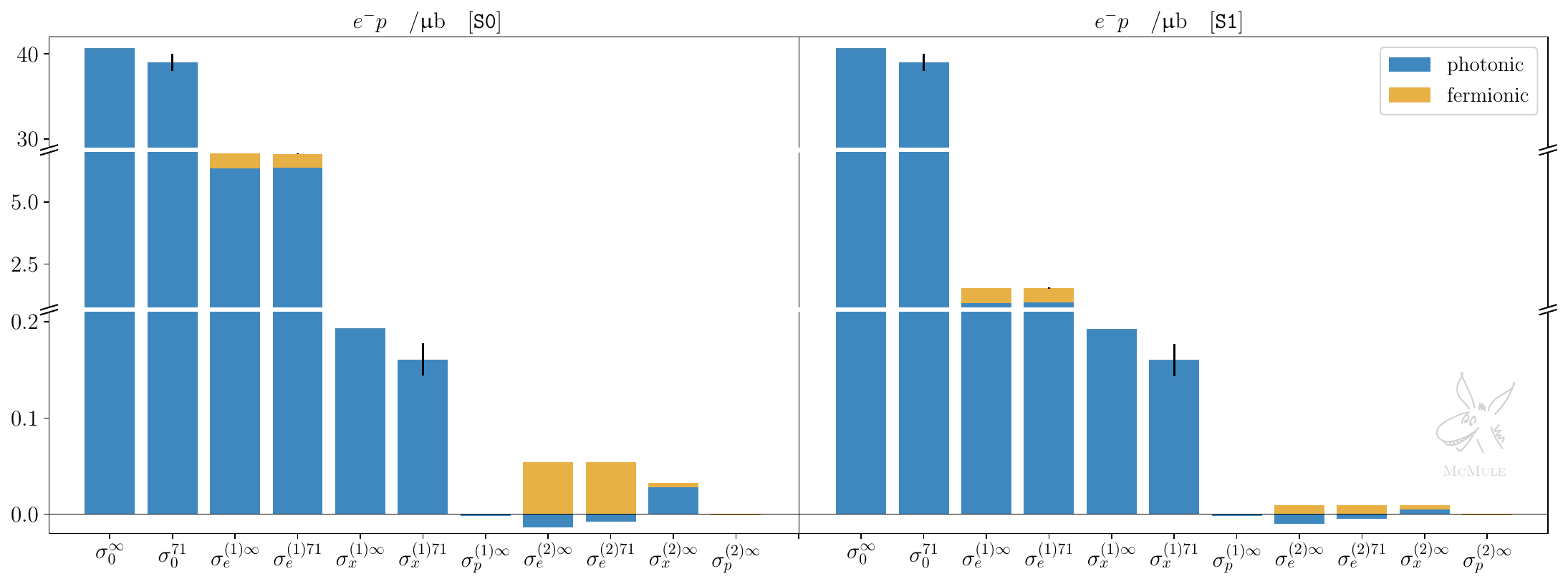}
  \caption{ 
  Integrated cross sections for $ep$ scattering, for both
    {\tt S0}~(left panel) and {\tt S1}~(right panel), at LO, NLO, and
    NNLO. Yellow bars indicate contributions with fermionic loop
    insertions, blue bars indicate photonic
    contributions. The fermionic contributions $\sigma_{\Pi}^{(1)\Lambda}$, $\sigma_{e\Pi}^{(2)\Lambda}$, $\sigma_{x\Pi}^{(2)\infty}$, and $\sigma_{p\Pi}^{(2)\infty}$ are individually combined with $\sigma_{e}^{(1)\Lambda}$, $\sigma_{e}^{(2)\Lambda}$, $\sigma_{x}^{(2)\infty}$, and $\sigma_{p}^{(2)\infty}$, respectively.} $\Lambda^2=0.71$ GeV$^2$ is taken as the reference value for the dipole parameter. Black bands denote variations due to
    $\Lambda^2\in [0.60,\, 0.86]\,{\rm GeV}^2$. Note that the black bands should not be interpreted as the uncertainty estimate of our theory prediction. They merely illustrate the impact of the form factors and their possible uncertainties, assuming our naive dipole ansatz. Further uncertainties, e.g., due to higher-order corrections, model dependence or missing inelastic TPE, are not shown.

  \label{fig:xsec-ep}
  \vspace{3em}
  \centering
  \includegraphics[width=1.\textwidth]{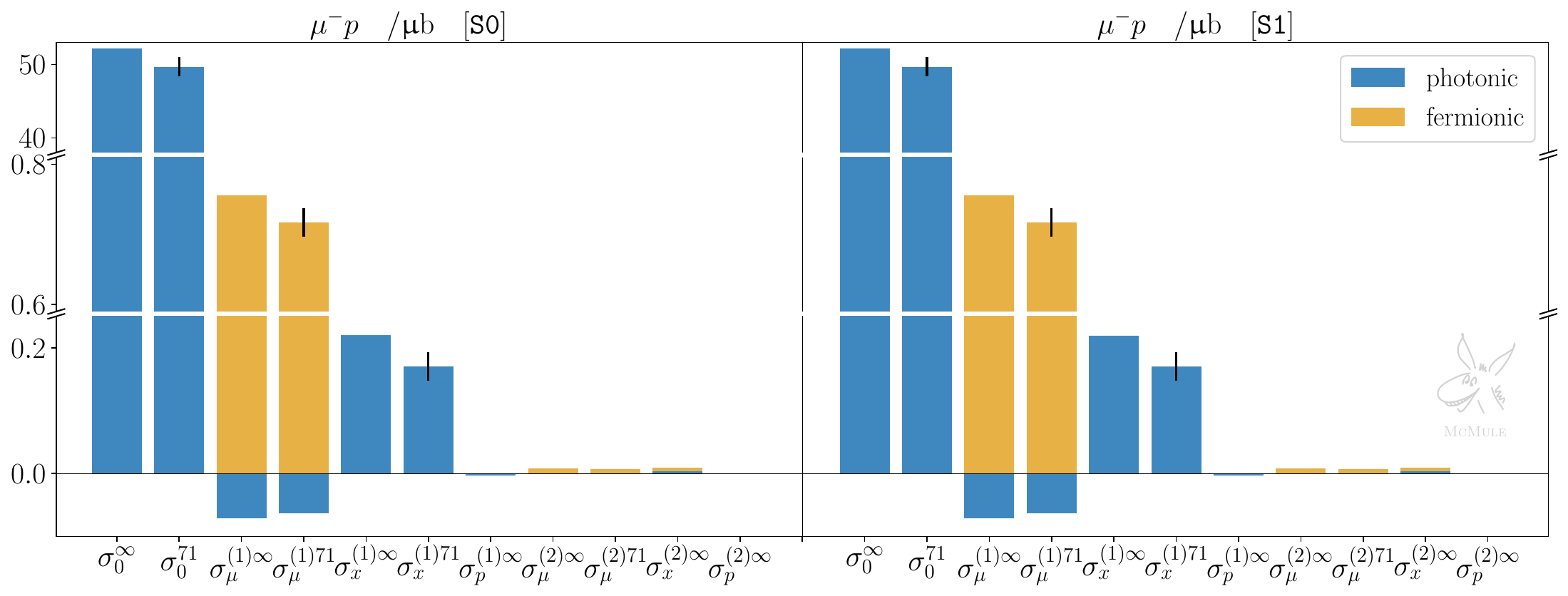}
  \caption{Same as Figure~\ref{fig:xsec-ep} but for $\mu p$ scattering.}
  \label{fig:xsec-mp}
\end{figure}

In order to better grasp the impact of different contributions to $e p$ and $\mu p$ scattering, Figures~\ref{fig:xsec-ep}
and~\ref{fig:xsec-mp} show the same corrections presented in
Tables~\ref{tab:tot_xsec_e} and~\ref{tab:tot_xsec_m} as bar plots, for
both kinematical scenarios. For each contribution, photonic and fermionic corrections are plotted with a different colour. Each contribution with a superscript ``71'' indicates the
corresponding pure QED contribution with the additional inclusion of proton from factors, using the dipole ansatz in (\ref{eq:lambda}) with
$\Lambda^2=0.71$ GeV$^2$. For the LO and NLO contributions, a black band
represents the variation obtained with $0.60\,{\rm GeV}^2 < \Lambda^2
< 0.86\,{\rm GeV}^2$. When a black band refers to a sum of photonic and fermionic corrections, this is plotted as the square root of the quadrature sum of the two contributions. In the case of $ep$ scattering, the black bands for the NLO electronic correction are particularly small compared to the scale of the plot, for both {\tt S0} and {\tt S1}.

The discussion about $ep$ and $\mu p$ scattering somewhat differs because of the different lepton mass. We start our
analysis with the former.
As touched upon in Section~\ref{sec:calc}, collinear photon emission introduces logarithms of the form $\log(m_i^2/E^2)$ where $m_i \in \{m_e,M\}$ depending on whether the photon is emitted from the electron or the proton line. In the former case, the logarithm is large and the corresponding contribution enhanced. This explains the hierarchy between the different photonic NLO contributions in Table~\ref{tab:tot_xsec_e}: the electronic OPE corrections are dominant compared to TPE, and even more compared to OPE protonic corrections. If a cut that limits hard forward-angle radiation ({\tt S1}) is applied, the collinear enhancement is reduced and the hierarchy is less pronounced.

In order to illustrate the relative importance of various corrections in $e^- p$ scattering w.r.t.~the TPE, 
we will use $\Lambda^2=0.71$\,GeV$^2$ for the reference form factor and
 $\sigma_{x}^{(1){71}}$ as normalisation. Electronic and fermionic NLO corrections by far outweigh the TPE,
\begin{equation}
   \frac{\{\sigma_e^{(1)71}, \sigma_\Pi^{(1)71}, |\sigma_p^{(1)}|\}}{\sigma_{x}^{(1){71}}} \approx
  \begin{cases}
    \{40, 3.5, 0.01\} & [{\tt S0}]\\
    \{\phantom{4}6, 3.5, 0.01\} & [{\tt S1}]
  \end{cases} \,, \quad   
  \label{comp:e2}
\end{equation}
even in the kinematical scenario {\tt S1} in which the electronic contribution is reduced by almost a factor 7 due to the cut on $E_\gamma$. 
Protonic corrections, on the other hand, are much smaller, justifying their neglect or approximate (pointlike proton) inclusion. 

The impact of form factor insertions at LO and NLO can be quantified, for both kinematical
scenarios, as  
\begin{align}
\label{comp:e1}
\begin{aligned}
  &\frac{|\sigma_0^{\infty}-\sigma_0^{71}|}{\sigma_{x}^{(1){71}}}
  \approx 10 \,, \quad \\
  &\frac{\{|\sigma_e^{(1)\infty}-\sigma_e^{(1){71}}|, 
  |\sigma_x^{(1)\infty}-\sigma_x^{(1){71}}|, 
  |\sigma_\Pi^{(1)\infty}-\sigma_\Pi^{(1){71}}|\}}{\sigma_{x}^{(1){71}}}
  \approx
  \begin{cases}
    \{0.06, 0.20, 0.15\} & [{\tt S0}]\\
    \{0.15  , 0.20, 0.15 \} & [{\tt S1}]
  \end{cases} \, .
\end{aligned}  
\end{align}
The first relation simply states that the impact of the form factor at tree level is 
clearly dominating any TPE effect, as expected. Comparing the other relations in 
\eqref{comp:e1} we note that the form factor insertion turns out to be more relevant for the mixed and fermionic corrections and less relevant for the electronic OPE correction.  With the same normalisation, the impact of varying
$\Lambda$ at LO and NLO can be quantified as
\begin{align}
\label{comp:e3}
\begin{aligned}
  &\frac{|\sigma_0^{86}-\sigma_0^{60}|}{\sigma_{x}^{(1){71}}}
  \approx 6 \,, \quad \\
  &\frac{\{|\sigma_e^{(1)86}-\sigma_e^{(1){60}}|, 
  |\sigma_x^{(1)86}-\sigma_x^{(1){60}}|, 
  |\sigma_\Pi^{(1)86}-\sigma_\Pi^{(1){60}}|\}}{\sigma_{x}^{(1){71}}}
  \approx
  \begin{cases}
    \{0.02, 0.11, 0.10\} & [{\tt S0}]\\
    \{0.08, 0.11, 0.10\} & [{\tt S1}]
  \end{cases} \, .
\end{aligned}  
\end{align}
Not surprisingly, variations in $\Lambda$ have the largest impact on the LO result. At NLO, the effect on the electronic, mixed, and fermionic corrections is roughly the same, and still more than half the size of the effect of the form factor inclusion itself, shown in \eqref{comp:e1}. Thus, in the context of $ep$ scattering, a
calculation of NLO corrections necessarily requires a precise inclusion of the proton form factors.

Comparing the NNLO pure QED corrections (involving also pointlike three-photon exchanges) to the TPE effects we find 
\begin{align}
\label{comp:e4}
\begin{aligned}
  & \frac{\{
    |\sigma_e^{(2)\infty}|,
    \sigma_{x}^{(2)\infty},
    \sigma_{e\Pi}^{(2)\infty},
    \sigma_{x\Pi}^{(2)\infty}
    \}}{\sigma_{x}^{(1){71}}}
  \approx
  \begin{cases}
    \{0.08, 0.17, 0.34, 0.03\} & [{\tt S0}]\\
    \{0.06, 0.03, 0.06, 0.03\} & [{\tt S1}]
  \end{cases} \,.
\end{aligned}  
\end{align}
Their relative size is of the same order (if not bigger) as the impact of adding form factors to the NLO TPE corrections, (\ref{comp:e1}), or considering uncertainties of the TPE implementation, (\ref{comp:e3}).  This
is particularly the case if forward energetic photons are not
restricted. Hence, a detailed effort to improve the description of TPE contributions needs to be combined with NNLO QED corrections.

The case of $\mu p$ scattering behaves differently in some respects. The
difference between muon and proton mass is much smaller than the
difference between electron and proton mass. This is why we do not
observe any collinear enhancements, and the NLO muonic corrections are  smaller than the electronic corrections in~\eqref{comp:e2},
\begin{equation}
   \frac{\{|\sigma_\mu^{(1)71}|, \sigma_\Pi^{(1)71}, |\sigma_p^{(1)}|\}}{\sigma_{x}^{(1){71}}} \approx
    \{0.4, 4.2,0 .02\} \,. \quad   
  \label{comp:m2}
\end{equation}
Again, the protonic corrections are small enough to justify their approximate (pointlike proton) inclusion. Here and in the following, we only discuss the scenario {\tt{S0}}, because the cut on energetic~(forward)
photons has a marginal effect only.

Considering the same quantities as for $e^- p$ scattering, \eqref{comp:e1}, but now for $\mu^- p$ scattering, again normalising by the TPE corrections $\sigma_{x}^{(1){71}}$, we find
\begin{align}
\label{comp:m1}
\hspace{-0.3cm}  \frac{|\sigma_0^{\infty}-\sigma_0^{71}|}{\sigma_{x}^{(1){71}}}
  \approx 15 \,, \quad
  \frac{\{|\sigma_\mu^{(1)\infty}-\sigma_\mu^{(1){71}}|, 
  |\sigma_x^{(1)\infty}-\sigma_x^{(1){71}}|, 
  |\sigma_\Pi^{(1)\infty}-\sigma_\Pi^{(1){71}}|\}}{\sigma_{x}^{(1){71}}}
  \approx
    \{0.05, 0.30, 0.20\} \, .
\end{align}
Thus, the impact of adding the form factor is slightly larger in the $\mu p$ case. Proceeding in analogy with the $e p$ case, we next consider the impact of varying $\Lambda$ for $\mu p$ 
\begin{equation}
  \label{comp:m3}
  \frac{|\sigma_0^{86}-\sigma_0^{60}|}{\sigma_{x}^{(1){71}}}
  \approx 8 \,, \quad
  \frac{\{|\sigma_\mu^{(1){86}}-\sigma_\mu^{(1){60}}|,
  |\sigma_{x}^{(1){86}}-\sigma_{x}^{(1){60}}|,
 |\sigma_{\Pi}^{(1){86}}-\sigma_{\Pi}^{(1){60}}| \} }{\sigma_{x}^{(1){71}}}
  \approx \{0.02,0.13,0.12\} 
\end{equation}
and note that the results are similar as in \eqref{comp:e3}. Finally, comparing TPE to pure NNLO QED, we get
\begin{align}
\label{comp:m4}
\begin{aligned}
  & \frac{\{
    |\sigma_\mu^{(2)\infty}|,
    \sigma_{x}^{(2)\infty},
    \sigma_{\mu\Pi}^{(2)\infty},
    \sigma_{x\Pi}^{(2)\infty}
    \}}{\sigma_{x}^{(1){71}}}
  \approx
    \{0.00, 0.02, 0.05, 0.03\}  \,.
\end{aligned}  
\end{align}
Here we see a clear difference between \eqref{comp:e4} and \eqref{comp:m4}. As expected, higher-order QED radiative corrections are less relevant for $\mu p$ scattering. The impact of variation in the TPE evaluation is larger than the pure NNLO QED corrections. From this perspective, it is thus advantageous to study TPE in $\mu p$ scattering. However, the pure NNLO QED corrections add up to 10\% of the TPE. Therefore, a precision study still benefits from inclusion of state-of-the-art QED corrections. 

\begin{figure}
  \centering
  \begin{minipage}{.49\textwidth}
    \centering
    \includegraphics[width=1.\textwidth]{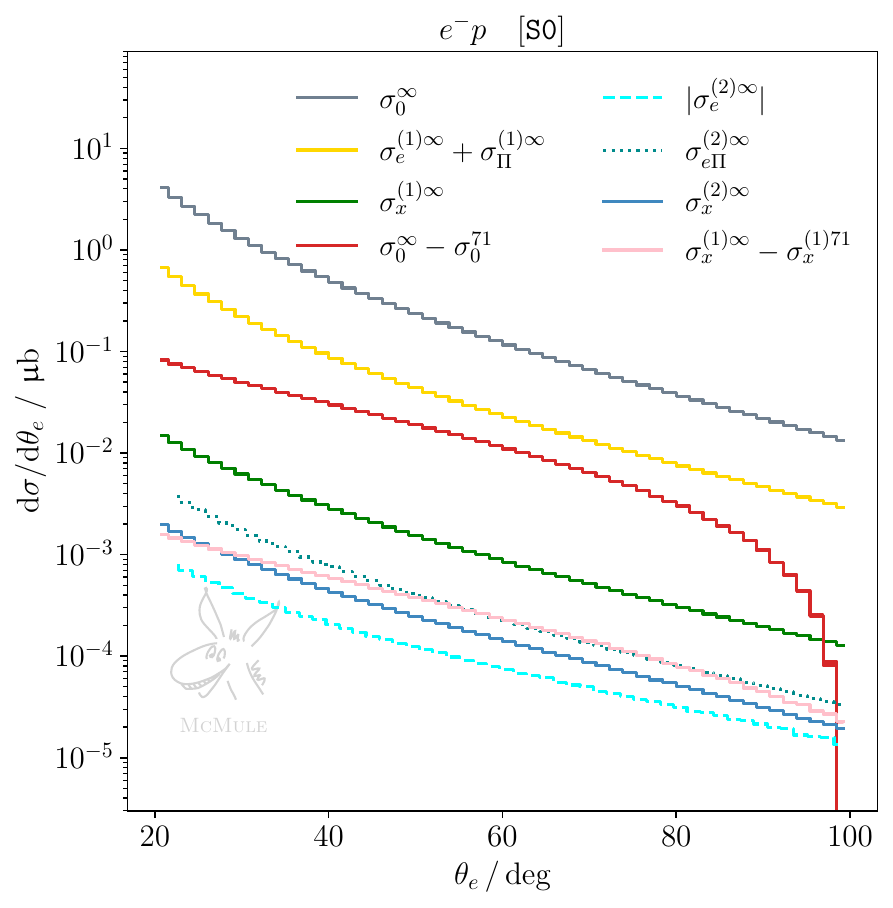}
  \end{minipage}%
  \begin{minipage}{.49\textwidth}
    \centering
    \includegraphics[width=1.\textwidth]{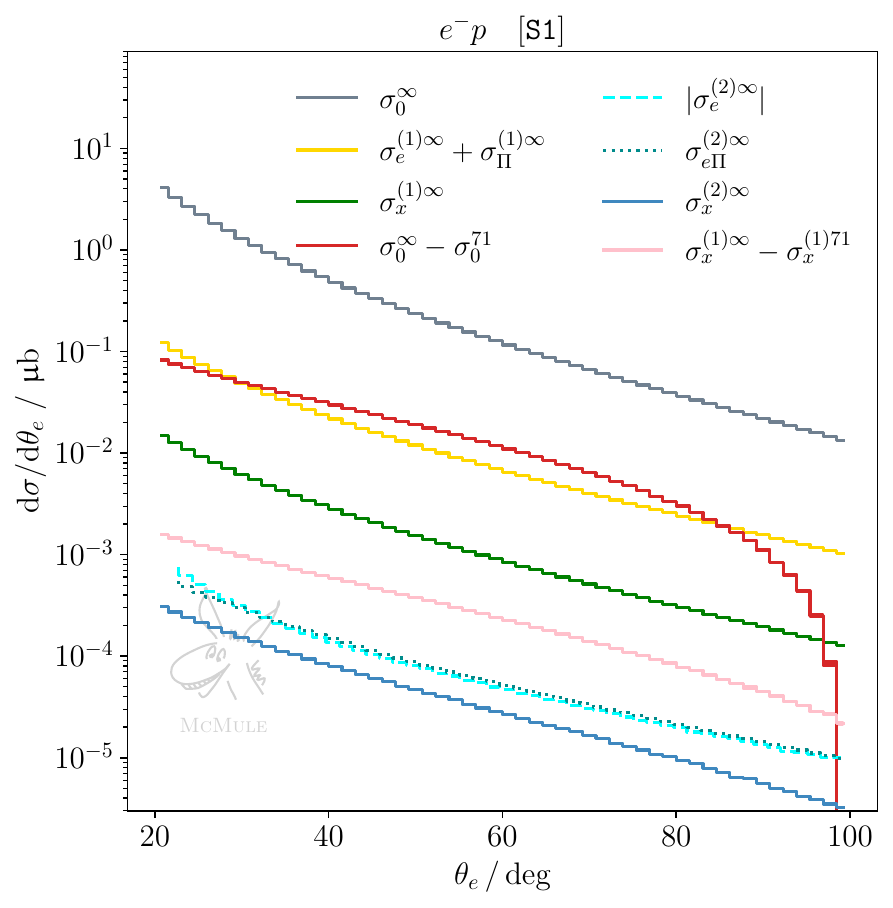}
  \end{minipage}
  \caption{Differential cross section for $ep$ scattering w.r.t.\ with respect to
    the electron scattering angle, for {\tt S0}~(left panel) and {\tt
      S1}~(right panel). The cross section is split into different
    contributions at LO~(gray), NLO~(yellow and green) and
    NNLO~(dashed cyan, dotted dark cyan and dark blue). The impact of
    the proton finite size on the OPE at LO is shown in red, and the impact on the elastic TPE at NLO is shown in pink. Some contributions are presented with their absolute value as the scale is logarithmic.}
  \label{fig:diff-ep}
\end{figure}

\begin{figure}
  \centering
  \begin{minipage}{.49\textwidth}
    \centering
    \includegraphics[width=1.\textwidth]{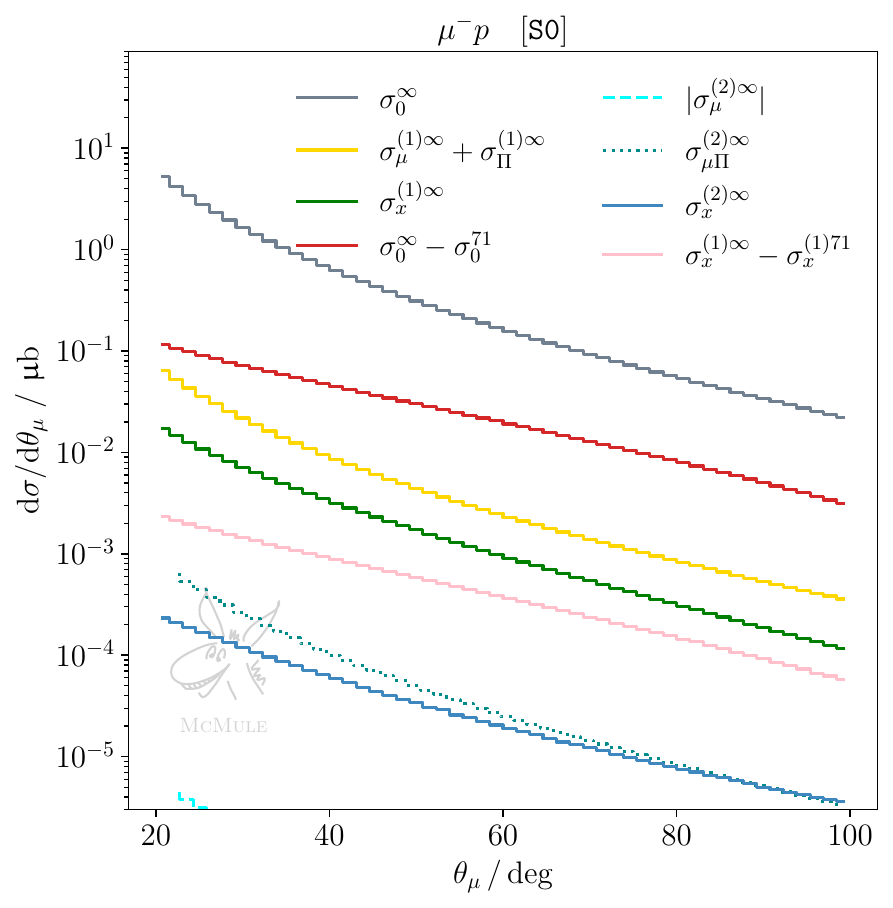}
  \end{minipage}%
  \begin{minipage}{.49\textwidth}
    \centering
    \includegraphics[width=1.\textwidth]{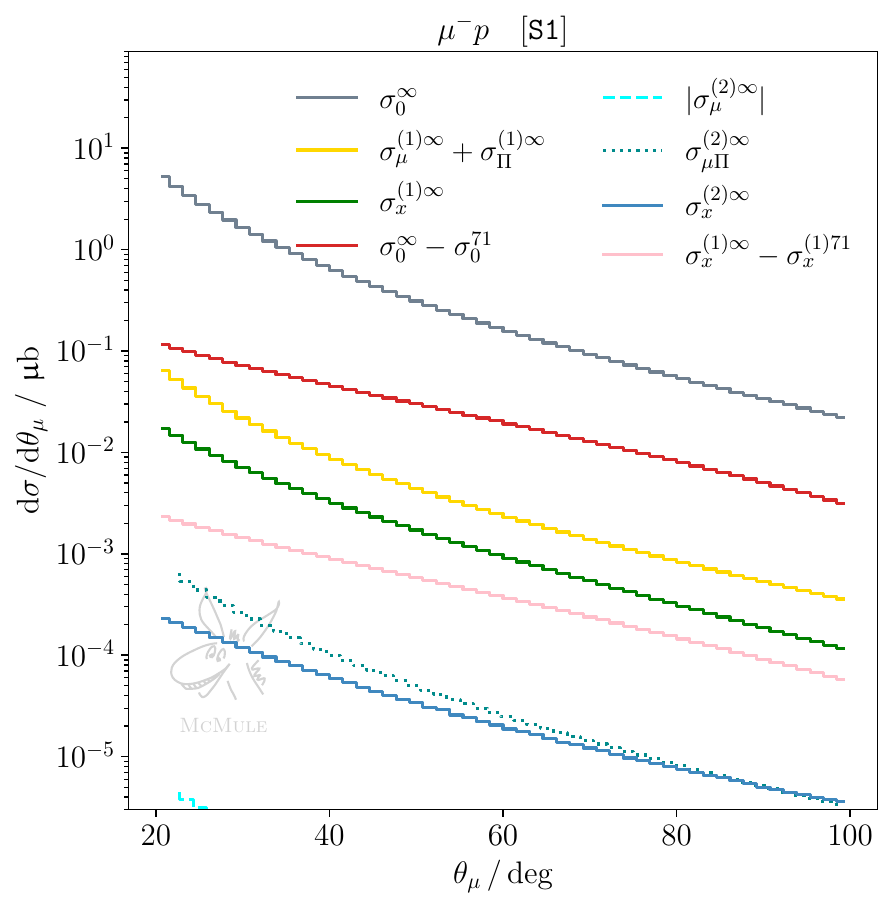}
  \end{minipage}
  \caption{Same as Figure~\ref{fig:diff-ep} but for $\mu p$ scattering.}
  \label{fig:diff-mp}
\end{figure}

We complement our discussion with results at differential level,
considering differential distributions w.r.t.\ the lepton
scattering angle. Figures~\ref{fig:diff-ep} and~\ref{fig:diff-mp}
present such distributions for both kinematical scenarios, for $ep$ and $\mu p$ scattering, respectively. In each plot, pure NLO leptonic and fermionic (yellow curve) as well as mixed corrections (green curve) are compared to the difference of LO effects with and without inclusion of the proton form factors (red curve). Furthermore, pure NNLO QED corrections (blue curves) are compared to the difference of NLO corrections with and without inclusion of the proton form factors (pink curve). Thus, the impact of the proton form factor inclusion at LO and NLO can be contrasted at differential level to NLO and NNLO pointlike corrections, respectively. 
The statements previously made for the integrated cross section are confirmed at the differential level. In the case of $ep$ scattering, NLO pure QED corrections outweigh the effect of the form factor inclusion at LO, and NNLO pure QED corrections are comparable to the impact of the form factor inclusion in the NLO TPE correction, particularly for {\tt S0}. The impact of pure QED corrections on $\mu p$ scattering, which can be read from the same set of curves, is smaller than in the $e p$ case but still not negligible.

\begin{figure}
  \centering
  \begin{minipage}{.49\textwidth}
    \centering
    \includegraphics[width=1.\textwidth]{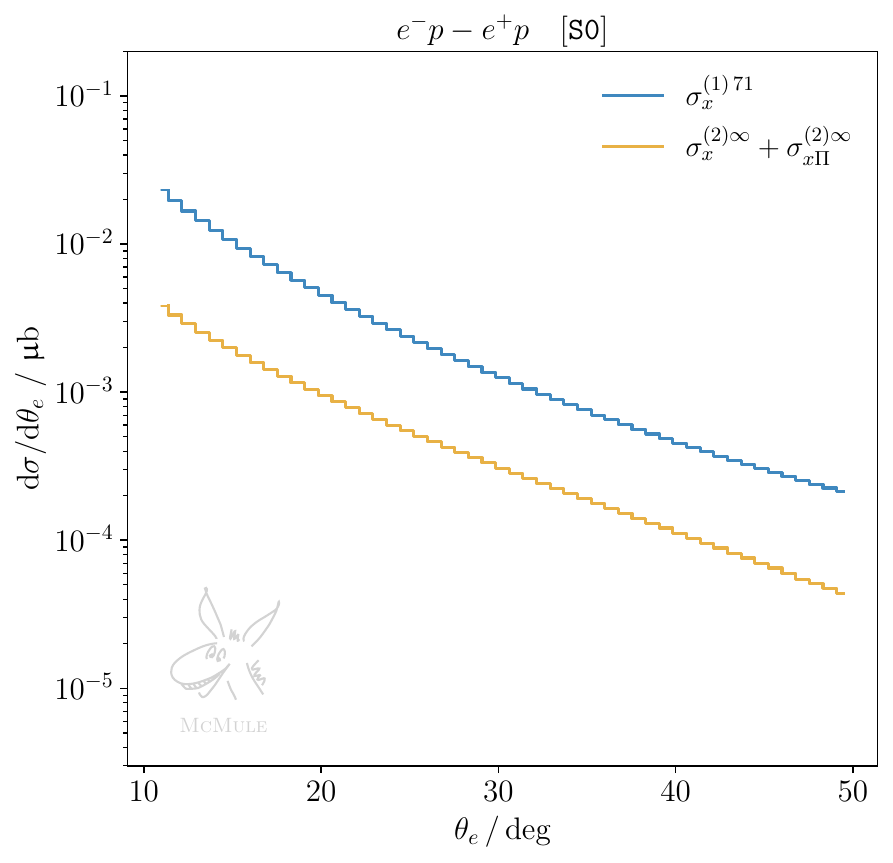}
  \end{minipage}
  \begin{minipage}{.49\textwidth}
    \centering
    \includegraphics[width=1.\textwidth]{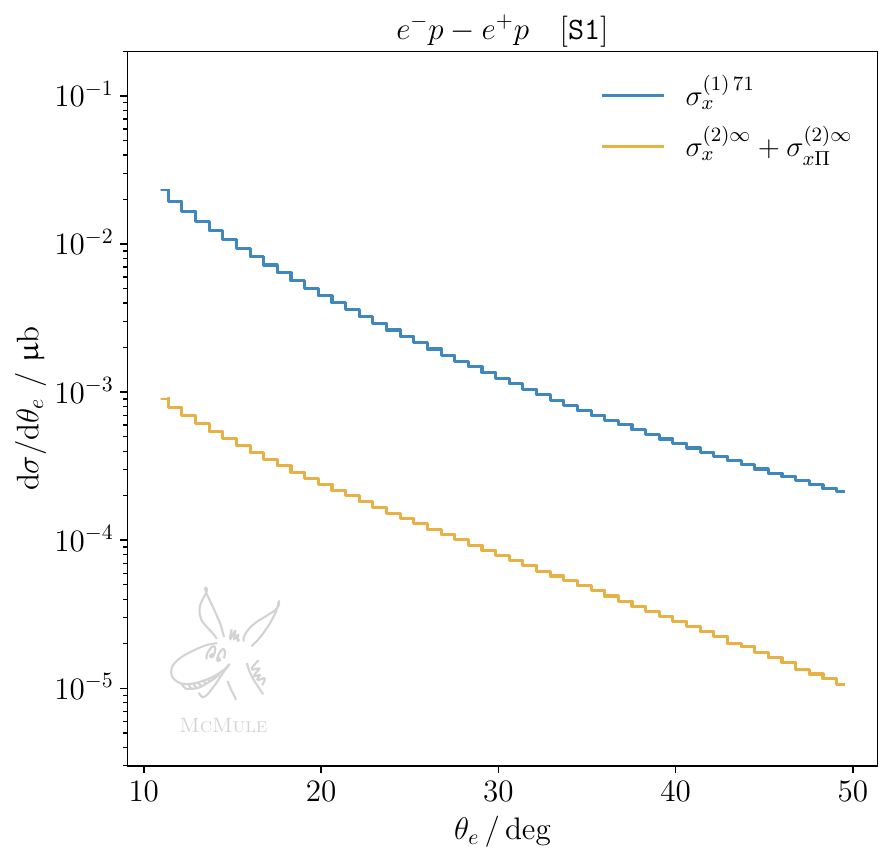}
  \end{minipage}
  \caption{Difference between the $\theta_e$ differential cross
    sections for $e^-p$ and $e^+p$ scattering, for {\tt S0}~(left
    panel) and {\tt S1}~(right panel). The corrections to the cross
    section are split into different contributions at NLO~(blue) and
    NNLO~(yellow). The only non-zero corrections are those with odd
    powers of the formal charge $q_\ell$.}
  \label{fig:diffx-ep}

\vspace{3em}
  \centering
  \begin{minipage}{.49\textwidth}
    \centering
    \includegraphics[width=1.\textwidth]{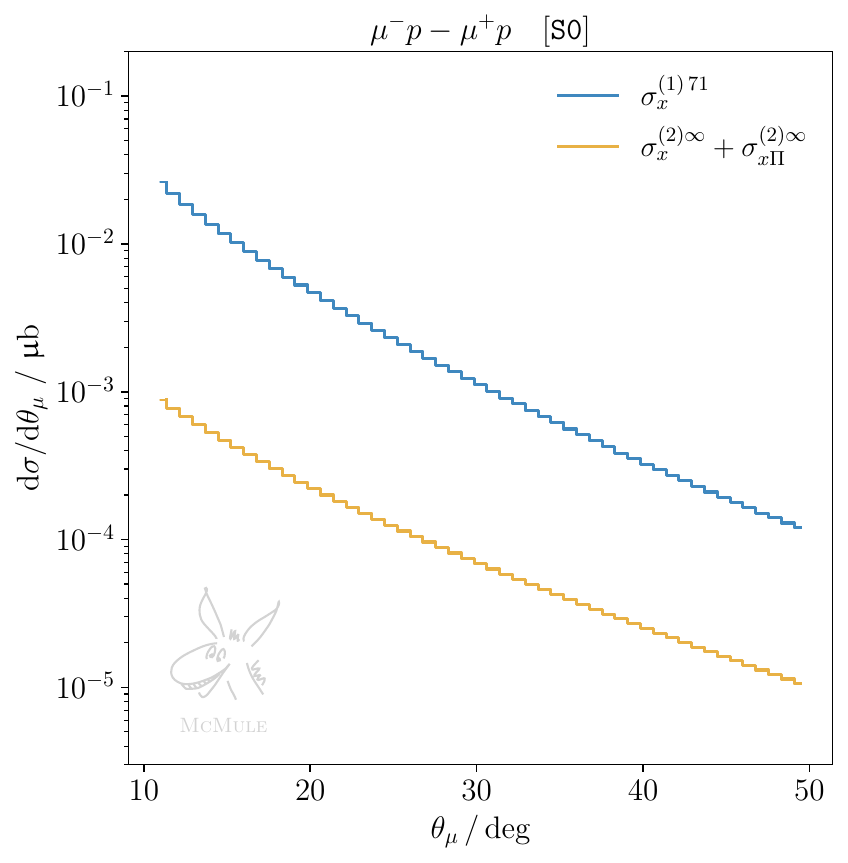}
  \end{minipage}
  \begin{minipage}{.49\textwidth}
    \centering
    \includegraphics[width=1.\textwidth]{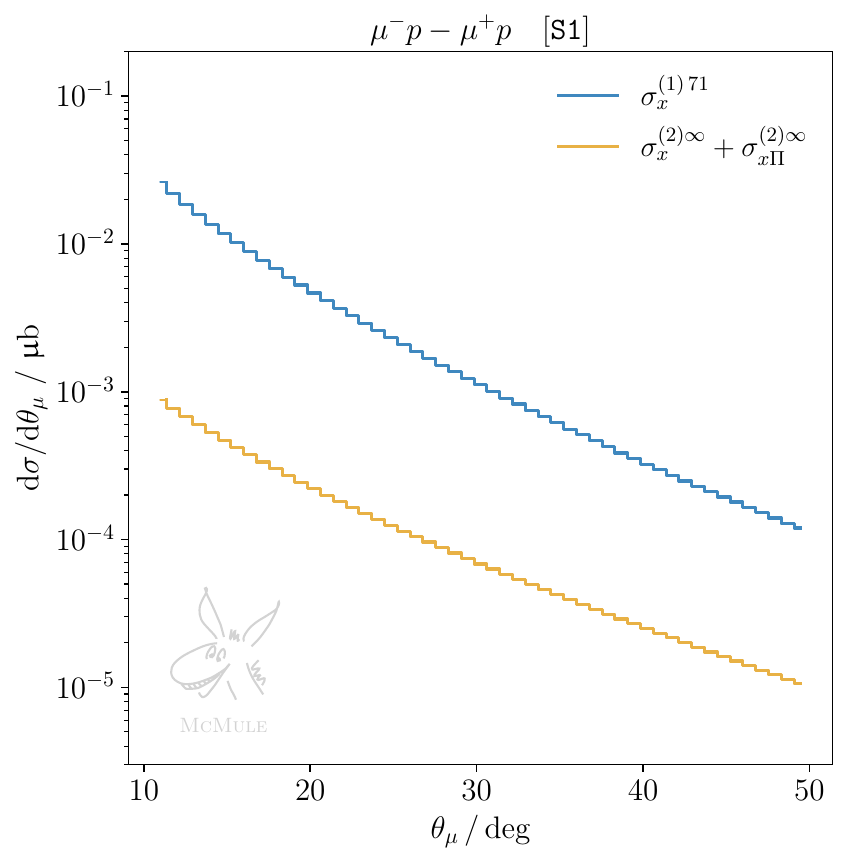}
  \end{minipage}
  \caption{Same as Figure~\ref{fig:diffx-ep} but for $\mu p$ scattering.}
  \label{fig:diffx-mp}
\end{figure}

The cross-section difference between $\ell^-p$ and $\ell^+p$ allows to cancel radiative contributions with an even power of the formal charge $q_\ell$. Thus, up to and including NNLO, this leaves only the $\sigma_x^{(1)}$ and $\sigma_{x\Pi}^{(2)}$  contributions, and a subset of the $\sigma_{x}^{(2)}$ contribution. Figures~\ref{fig:diffx-ep} and \ref{fig:diffx-mp} show the latter in
both kinematical scenarios for $ep$ and $\mu p$ scattering, respectively. One can see again that, in the case of {\tt S1} for $ep$ scattering and in general for $\mu p$ scattering, NNLO corrections are more suppressed than in {\tt S0} for $ep$. Nevertheless, when extracting the TPE effect empirically from the cross-section difference measured at MUSE, i.e. in scenario {\tt S1} for $ep$ or $\mu p$ scattering, it is important to take higher-order radiative corrections into account. In both cases, NNLO contributions will lead to a $10\%$ correction on the extraction, thus, cannot be neglected.

\section{Conclusions and outlook} \label{sec:concl}

We have presented an update of the \mcmule{} framework for the process of lepton-proton scattering \cite{Banerjee:2020rww} with inclusion of additional proton-structure effects from elastic TPE, and complete pointlike QED corrections at NNLO, with lepton-mass effects. In Section~\ref{sec:calc}, we have given a detailed description of the contributions that are included in the latest version of \mcmule{}. Our notation for individual contributions has been introduced in (\ref{eg:xsnnlo}) -- (\ref{eq:rescompact}). In Section~\ref{sec:res}, we have studied the  impact of higher-order QED
radiative corrections on the unpolarised cross section for lepton-proton scattering at MUSE, focusing on one particular choice of beam momentum ($p_\text{beam}=210$\,MeV). The availability of
both electrons and muons, with both polarities, is a remarkable
advantage for the MUSE experiment, as it allows to analyse a diversified
phenomenology and to keep under control QED radiative corrections, if
needed. This is achieved either with physical cuts on hard forward
photons or by using muons, which are less inclined to
irradiate.

Hadronic corrections are usually known less precisely than pure QED corrections (with a pointlike proton). In this work,
our main aim has been to  assess the relative size of NNLO pure QED corrections, as compared to the LO and NLO corrections with inclusion of the proton form factors and their uncertainties. A particular focus has been on TPE effects, referred to also as the NLO mixed corrections.  Since the MUSE kinematics is limited to the low momentum-transfer region ($Q^2<0.08$ GeV$^2$), the inelastic TPE is small enough to be neglected in view of the anticipated $1\%$ accuracy of the cross section measurement. Therefore, only the elastic TPE has been implemented through a simple hadronic model assuming on-shell proton form factors described by a dipole ansatz \eqref{eq:FF}. The dipole parameter \eqref{eq:lambda} has been varied around the standard dipole $\Lambda^2=0.71$ GeV$^2$ within a broad range $0.60\,{\rm GeV}^2 < \Lambda^2
< 0.86\,{\rm GeV}^2$ to illustrate the impact of form factors uncertainties.

We conclude that while it is sufficient to evaluate the protonic NLO corrections in pure QED with a pointlike proton, all other NLO corrections, in particular the mixed and fermionic, necessarily require a precise inclusion of the proton form factors. Furthermore, we haven shown that NNLO pure QED corrections can be almost as sizeable as the NLO TPE corrections. Even for $ep$ scattering with cuts on hard forward photon emission (scenario {\tt S1}), see \eqref{comp:e4}, or for $\mu p$ scattering, see \eqref{comp:m4}, where higher-order radiative corrections are more suppressed, NNLO QED corrections should always be included together with an improved  description of TPE effects. Equivalently, NNLO pure QED corrections need to be included when extracting the TPE effect empirically to better than $10\%$ accuracy from the cross-section difference between $\ell^-p$ and $\ell^+p$ scattering.

The same analysis can be readily
repeated within the \mcmule{} framework for different kinematical
scenarios or other observables, and also broadened to cover further
experiments with different $Q^2$ ranges. To this end,  the implementation of the elastic TPE correction with input from modern form-factor parametrisations, and the implementation of inelastic TPE corrections, are planned for a future version of \mcmule{}.

\section*{Acknowledgements}
It is a pleasure to thank E.~Cline and R.~Gilman for discussions regarding MUSE. We are grateful to J.~Bernauer, as well as O.~Tomalak and M.~Vanderhaeghen for providing their results from \cite{A1:2013fsc} and \cite{Tomalak:2015hva}, respectively. 
T.E. was supported by the German Federal Ministry for Education and Research (BMBF) under contract no. 05H21VFCAA.
F.H.~and V.S.~acknowledge support by the Swiss National Science Foundation (SNSF) through the Ambizione Grant PZ00P2\_193383. F.H.~acknowledges support by the Deutsche Forschungsgemeinschaft (DFG) through the Emmy Noether Programme (grant 449369623) and the Research Unit FOR 5327 “Photon-photon interactions in the Standard Model and beyond --- exploiting the discovery potential from MESA to the LHC” (grant 458854507). 
M.R. is supported by the SNSF Grant 200020\_207386.  
YU acknowledges support by the UK Science and Technology Facilities Council (STFC) under grant ST/T001011/1.
\bibliographystyle{JHEP}
\bibliography{megmule}{}

\providecommand{\href}[2]{#2}\begingroup\raggedright\begin{thebibliography}{10}

\bibitem{Afanasev:2023gev}
A.~Afanasev et~al., \emph{{Radiative Corrections: From Medium to High Energy
  Experiments}},  \href{https://arxiv.org/abs/2306.14578}{{\ttfamily
  2306.14578}}.

\bibitem{Mo:1968cg}
L.~W. Mo and Y.-S. Tsai, \emph{{Radiative Corrections to Elastic and Inelastic
  $e p$ and $\mu p$ Scattering}},
  \href{https://doi.org/10.1103/RevModPhys.41.205}{\emph{Rev. Mod. Phys.}
  {\bfseries 41} (1969) 205}.

\bibitem{Maximon:2000hm}
L.~C. Maximon and J.~A. Tjon, \emph{{Radiative corrections to electron proton
  scattering}}, \href{https://doi.org/10.1103/PhysRevC.62.054320}{\emph{Phys.
  Rev. C} {\bfseries 62} (2000) 054320}.

\bibitem{Bystritskiy:2007hw}
Y.~Bystritskiy, E.~Kuraev and E.~Tomasi-Gustafsson, \emph{{Structure function
  method applied to polarized and unpolarized electron-proton scattering: A
  solution of the $G_E(p)$/$G_M(p)$ discrepancy}},
  \href{https://doi.org/10.1103/PhysRevC.75.015207}{\emph{Phys. Rev.}
  {\bfseries C75} (2007) 015207}.

\bibitem{Kuraev:2013dra}
E.~A. Kuraev, A.~I. Ahmadov, Y.~M. Bystritskiy and E.~Tomasi-Gustafsson,
  \emph{{Radiative corrections for electron proton elastic scattering taking
  into account high orders and hard photon emission}},
  \href{https://doi.org/10.1103/PhysRevC.89.065207}{\emph{Phys. Rev. C}
  {\bfseries 89} (2014) 065207}
  [\href{https://arxiv.org/abs/1311.0370}{{\ttfamily 1311.0370}}].

\bibitem{Gramolin:2014pva}
A.~V. Gramolin, V.~S. Fadin, A.~L. Feldman, R.~E. Gerasimov, D.~M. Nikolenko,
  I.~A. Rachek et~al., \emph{{A new event generator for the elastic scattering
  of charged leptons on protons}},
  \href{https://doi.org/10.1088/0954-3899/41/11/115001}{\emph{J. Phys.}
  {\bfseries G41} (2014) 115001}.

\bibitem{Gerasimov:2015aoa}
R.~E. Gerasimov and V.~S. Fadin, \emph{{Analysis of approximations used in
  calculations of radiative corrections to electron-proton scattering cross
  section}}, \href{https://doi.org/10.1134/S1063778815010081}{\emph{Phys. Atom.
  Nucl.} {\bfseries 78} (2015) 69}.

\bibitem{Akushevich:2015toa}
I.~Akushevich, H.~Gao, A.~Ilyichev and M.~Meziane, \emph{{Radiative corrections
  beyond the ultra relativistic limit in unpolarized ep elastic and M\o{}ller
  scatterings for the PRad Experiment at Jefferson Laboratory}},
  \href{https://doi.org/10.1140/epja/i2015-15001-8}{\emph{Eur. Phys. J. A}
  {\bfseries 51} (2015) 1}.

\bibitem{Bucoveanu:2018soy}
R.~D. Bucoveanu and H.~Spiesberger, \emph{{Second-Order Leptonic Radiative
  Corrections for Lepton-Proton Scattering}},
  \href{https://doi.org/10.1140/epja/i2019-12727-1}{\emph{Eur. Phys. J. A}
  {\bfseries 55} (2019) 57} [\href{https://arxiv.org/abs/1811.04970}{{\ttfamily
  1811.04970}}].

\bibitem{Banerjee:2020rww}
P.~Banerjee, T.~Engel, A.~Signer and Y.~Ulrich, \emph{{QED at NNLO with {\sc
  McMule}}}, \href{https://doi.org/10.21468/SciPostPhys.9.2.027}{\emph{SciPost
  Phys.} {\bfseries 9} (2020) 027}
  [\href{https://arxiv.org/abs/2007.01654}{{\ttfamily 2007.01654}}].

\bibitem{CarloniCalame:2020yoz}
C.~M. Carloni~Calame, M.~Chiesa, S.~M. Hasan, G.~Montagna, O.~Nicrosini and
  F.~Piccinini, \emph{{Towards muon-electron scattering at NNLO}},
  \href{https://doi.org/10.1007/JHEP11(2020)028}{\emph{JHEP} {\bfseries 11}
  (2020) 028} [\href{https://arxiv.org/abs/2007.01586}{{\ttfamily
  2007.01586}}].

\bibitem{Banerjee:2021mty}
P.~Banerjee, T.~Engel, N.~Schalch, A.~Signer and Y.~Ulrich, \emph{{Bhabha
  scattering at NNLO with next-to-soft stabilisation}},
  \href{https://doi.org/10.1016/j.physletb.2021.136547}{\emph{Phys. Lett. B}
  {\bfseries 820} (2021) 136547}
  [\href{https://arxiv.org/abs/2106.07469}{{\ttfamily 2106.07469}}].

\bibitem{Banerjee:2021qvi}
P.~Banerjee, T.~Engel, N.~Schalch, A.~Signer and Y.~Ulrich, \emph{{M{\o}ller
  scattering at NNLO}},
  \href{https://doi.org/10.1103/PhysRevD.105.L031904}{\emph{Phys. Rev. D}
  {\bfseries 105} (2022) L031904}
  [\href{https://arxiv.org/abs/2107.12311}{{\ttfamily 2107.12311}}].

\bibitem{Broggio:2022htr}
A.~Broggio et~al., \emph{{Muon-electron scattering at NNLO}},
  \href{https://doi.org/10.1007/JHEP01(2023)112}{\emph{JHEP} {\bfseries 01}
  (2023) 112} [\href{https://arxiv.org/abs/2212.06481}{{\ttfamily
  2212.06481}}].

\bibitem{Kollatzsch:2022bqa}
S.~Kollatzsch and Y.~Ulrich, \emph{{Lepton pair production at NNLO in QED with
  EW effects}},  \href{https://arxiv.org/abs/2210.17172}{{\ttfamily
  2210.17172}}.

\bibitem{A1:2013fsc}
{\scshape A1} collaboration, J.~C. Bernauer et~al., \emph{{Electric and
  magnetic form factors of the proton}},
  \href{https://doi.org/10.1103/PhysRevC.90.015206}{\emph{Phys. Rev. C}
  {\bfseries 90} (2014) 015206}
  [\href{https://arxiv.org/abs/1307.6227}{{\ttfamily 1307.6227}}].

\bibitem{Mihovilovic:2016rkr}
M.~Mihovilovi\v{c} et~al., \emph{{First measurement of proton's charge form
  factor at very low $Q^2$ with initial state radiation}},
  \href{https://doi.org/10.1016/j.physletb.2017.05.031}{\emph{Phys. Lett. B}
  {\bfseries 771} (2017) 194}
  [\href{https://arxiv.org/abs/1612.06707}{{\ttfamily 1612.06707}}].

\bibitem{Mihovilovic:2019jiz}
M.~Mihovilovi\v{c} et~al., \emph{{The proton charge radius extracted from the
  initial-state radiation experiment at MAMI}},
  \href{https://doi.org/10.1140/epja/s10050-021-00414-x}{\emph{Eur. Phys. J. A}
  {\bfseries 57} (2021) 107}
  [\href{https://arxiv.org/abs/1905.11182}{{\ttfamily 1905.11182}}].

\bibitem{Xiong:2019umf}
W.~Xiong et~al., \emph{{A small proton charge radius from an electron–proton
  scattering experiment}},
  \href{https://doi.org/10.1038/s41586-019-1721-2}{\emph{Nature} {\bfseries
  575} (2019) 147}.

\bibitem{Vanderhaeghen:2000ws}
M.~Vanderhaeghen, J.~M. Friedrich, D.~Lhuillier, D.~Marchand, L.~Van~Hoorebeke
  and J.~Van~de Wiele, \emph{{QED radiative corrections to virtual Compton
  scattering}}, \href{https://doi.org/10.1103/PhysRevC.62.025501}{\emph{Phys.
  Rev. C} {\bfseries 62} (2000) 025501}
  [\href{https://arxiv.org/abs/hep-ph/0001100}{{\ttfamily hep-ph/0001100}}].

\bibitem{Choudhary:2023rsz}
P.~Choudhary, U.~Raha, F.~Myhrer and D.~Chakrabarti, \emph{{Analytical
  Evaluation of Elastic Lepton-Proton Two-Photon Exchange in Chiral
  Perturbation Theory}},  \href{https://arxiv.org/abs/2306.09454}{{\ttfamily
  2306.09454}}.

\bibitem{Dye:2016uep}
S.~P. Dye, M.~Gonderinger and G.~Paz, \emph{{Elements of QED-NRQED effective
  field theory: NLO scattering at leading power}},
  \href{https://doi.org/10.1103/PhysRevD.94.013006}{\emph{Phys. Rev. D}
  {\bfseries 94} (2016) 013006}
  [\href{https://arxiv.org/abs/1602.07770}{{\ttfamily 1602.07770}}].

\bibitem{Dye:2018rgg}
S.~P. Dye, M.~Gonderinger and G.~Paz, \emph{{Elements of QED-NRQED Effective
  Field Theory: II. Matching of Contact Interactions}},
  \href{https://doi.org/10.1103/PhysRevD.100.054010}{\emph{Phys. Rev. D}
  {\bfseries 100} (2019) 054010}
  [\href{https://arxiv.org/abs/1812.05056}{{\ttfamily 1812.05056}}].

\bibitem{Tomalak:2016vbf}
O.~Tomalak, B.~Pasquini and M.~Vanderhaeghen, \emph{{Two-photon exchange
  corrections to elastic $e^-$-proton scattering: Full dispersive treatment of
  $\pi N$ states at low momentum transfers}},
  \href{https://doi.org/10.1103/PhysRevD.95.096001}{\emph{Phys. Rev. D}
  {\bfseries 95} (2017) 096001}
  [\href{https://arxiv.org/abs/1612.07726}{{\ttfamily 1612.07726}}].

\bibitem{Tomalak:2017shs}
O.~Tomalak, B.~Pasquini and M.~Vanderhaeghen, \emph{{Two-photon exchange
  contribution to elastic $e^-$ -proton scattering: Full dispersive treatment
  of $\pi$N states and comparison with data}},
  \href{https://doi.org/10.1103/PhysRevD.96.096001}{\emph{Phys. Rev. D}
  {\bfseries 96} (2017) 096001}
  [\href{https://arxiv.org/abs/1708.03303}{{\ttfamily 1708.03303}}].

\bibitem{Ahmed:2020uso}
J.~Ahmed, P.~G. Blunden and W.~Melnitchouk, \emph{{Two-photon exchange from
  intermediate state resonances in elastic electron-proton scattering}},
  \href{https://doi.org/10.1103/PhysRevC.102.045205}{\emph{Phys. Rev. C}
  {\bfseries 102} (2020) 045205}
  [\href{https://arxiv.org/abs/2006.12543}{{\ttfamily 2006.12543}}].

\bibitem{Guichon:2003qm}
P.~A.~M. Guichon and M.~Vanderhaeghen, \emph{{How to reconcile the Rosenbluth
  and the polarization transfer method in the measurement of the proton
  form-factors}},
  \href{https://doi.org/10.1103/PhysRevLett.91.142303}{\emph{Phys. Rev. Lett.}
  {\bfseries 91} (2003) 142303}
  [\href{https://arxiv.org/abs/hep-ph/0306007}{{\ttfamily hep-ph/0306007}}].

\bibitem{Blunden:2003sp}
P.~G. Blunden, W.~Melnitchouk and J.~A. Tjon, \emph{{Two photon exchange and
  elastic electron proton scattering}},
  \href{https://doi.org/10.1103/PhysRevLett.91.142304}{\emph{Phys. Rev. Lett.}
  {\bfseries 91} (2003) 142304}
  [\href{https://arxiv.org/abs/nucl-th/0306076}{{\ttfamily nucl-th/0306076}}].

\bibitem{Kondratyuk:2005kk}
S.~Kondratyuk, P.~G. Blunden, W.~Melnitchouk and J.~A. Tjon, \emph{{Delta
  resonance contribution to two-photon exchange in electron-proton
  scattering}},
  \href{https://doi.org/10.1103/PhysRevLett.95.172503}{\emph{Phys. Rev. Lett.}
  {\bfseries 95} (2005) 172503}
  [\href{https://arxiv.org/abs/nucl-th/0506026}{{\ttfamily nucl-th/0506026}}].

\bibitem{Chen:2004tw}
Y.~C. Chen, A.~Afanasev, S.~J. Brodsky, C.~E. Carlson and M.~Vanderhaeghen,
  \emph{{Partonic calculation of the two photon exchange contribution to
  elastic electron proton scattering at large momentum transfer}},
  \href{https://doi.org/10.1103/PhysRevLett.93.122301}{\emph{Phys. Rev. Lett.}
  {\bfseries 93} (2004) 122301}
  [\href{https://arxiv.org/abs/hep-ph/0403058}{{\ttfamily hep-ph/0403058}}].

\bibitem{Carlson:2007sp}
C.~E. Carlson and M.~Vanderhaeghen, \emph{{Two-Photon Physics in Hadronic
  Processes}},
  \href{https://doi.org/10.1146/annurev.nucl.57.090506.123116}{\emph{Ann. Rev.
  Nucl. Part. Sci.} {\bfseries 57} (2007) 171}
  [\href{https://arxiv.org/abs/hep-ph/0701272}{{\ttfamily hep-ph/0701272}}].

\bibitem{Arrington:2011dn}
J.~Arrington, P.~G. Blunden and W.~Melnitchouk, \emph{{Review of two-photon
  exchange in electron scattering}},
  \href{https://doi.org/10.1016/j.ppnp.2011.07.003}{\emph{Prog. Part. Nucl.
  Phys.} {\bfseries 66} (2011) 782}
  [\href{https://arxiv.org/abs/1105.0951}{{\ttfamily 1105.0951}}].

\bibitem{Afanasev:2017gsk}
A.~Afanasev, P.~G. Blunden, D.~Hasell and B.~A. Raue, \emph{{Two-photon
  exchange in elastic electron\textendash{}proton scattering}},
  \href{https://doi.org/10.1016/j.ppnp.2017.03.004}{\emph{Prog. Part. Nucl.
  Phys.} {\bfseries 95} (2017) 245}
  [\href{https://arxiv.org/abs/1703.03874}{{\ttfamily 1703.03874}}].

\bibitem{Borisyuk:2019gym}
D.~Borisyuk and A.~Kobushkin, \emph{{Two-Photon Exchange in Elastic Electron
  Scattering on Hadronic Systems}},
  \href{https://doi.org/10.15407/ujpe66.1.3}{\emph{Ukr. J. Phys.} {\bfseries
  66} (2021) 3} [\href{https://arxiv.org/abs/1911.10956}{{\ttfamily
  1911.10956}}].

\bibitem{MUSE:2017dod}
{\scshape MUSE} collaboration, R.~Gilman et~al., \emph{{Technical Design Report
  for the Paul Scherrer Institute Experiment R-12-01.1: Studying the Proton
  ''Radius'' Puzzle with $\mu p$ Elastic Scattering}},
  \href{https://arxiv.org/abs/1709.09753}{{\ttfamily 1709.09753}}.

\bibitem{Cline:2021ehf}
E.~Cline, J.~Bernauer, E.~J. Downie and R.~Gilman, \emph{{MUSE: The MUon
  Scattering Experiment}},
  \href{https://doi.org/10.21468/SciPostPhysProc.5.023}{\emph{SciPost Phys.
  Proc.} {\bfseries 5} (2021) 023}.

\bibitem{Li:2023sxf}
L.~Li et~al., \emph{{Instrumental uncertainties in radiative corrections for
  the MUSE experiment}},  \href{https://arxiv.org/abs/2307.06417}{{\ttfamily
  2307.06417}}.

\bibitem{Tomalak:2015hva}
O.~Tomalak and M.~Vanderhaeghen, \emph{{Two-photon exchange correction to
  muon\textendash{}proton elastic scattering at low momentum transfer}},
  \href{https://doi.org/10.1140/epjc/s10052-016-3966-3}{\emph{Eur. Phys. J. C}
  {\bfseries 76} (2016) 125}
  [\href{https://arxiv.org/abs/1512.09113}{{\ttfamily 1512.09113}}].

\bibitem{Engel:2019nfw}
T.~Engel, A.~Signer and Y.~Ulrich, \emph{{A subtraction scheme for massive
  QED}}, \href{https://doi.org/10.1007/JHEP01(2020)085}{\emph{JHEP} {\bfseries
  01} (2020) 085} [\href{https://arxiv.org/abs/1909.10244}{{\ttfamily
  1909.10244}}].

\bibitem{Borah:2020gte}
K.~Borah, R.~J. Hill, G.~Lee and O.~Tomalak, \emph{{Parametrization and
  applications of the low-$Q^2$ nucleon vector form factors}},
  \href{https://doi.org/10.1103/PhysRevD.102.074012}{\emph{Phys. Rev. D}
  {\bfseries 102} (2020) 074012}
  [\href{https://arxiv.org/abs/2003.13640}{{\ttfamily 2003.13640}}].

\bibitem{Sick:2012zz}
I.~Sick, \emph{{Problems with proton radii}},
  \href{https://doi.org/10.1016/j.ppnp.2012.01.013}{\emph{Prog. Part. Nucl.
  Phys.} {\bfseries 67} (2012) 473}.

\bibitem{Lin:2021xrc}
Y.-H. Lin, H.-W. Hammer and U.-G. Mei\ss{}ner, \emph{{New Insights into the
  Nucleon\textquoteright{}s Electromagnetic Structure}},
  \href{https://doi.org/10.1103/PhysRevLett.128.052002}{\emph{Phys. Rev. Lett.}
  {\bfseries 128} (2022) 052002}
  [\href{https://arxiv.org/abs/2109.12961}{{\ttfamily 2109.12961}}].

\bibitem{Antognini:2013txn}
A.~Antognini et~al., \emph{{Proton Structure from the Measurement of $2S-2P$
  Transition Frequencies of Muonic Hydrogen}},
  \href{https://doi.org/10.1126/science.1230016}{\emph{Science} {\bfseries 339}
  (2013) 417}.

\bibitem{Mohr:2015ccw}
P.~J. Mohr, D.~B. Newell and B.~N. Taylor, \emph{{CODATA Recommended Values of
  the Fundamental Physical Constants: 2014}},
  \href{https://doi.org/10.1103/RevModPhys.88.035009}{\emph{Rev. Mod. Phys.}
  {\bfseries 88} (2016) 035009}
  [\href{https://arxiv.org/abs/1507.07956}{{\ttfamily 1507.07956}}].

\bibitem{Tiesinga:2021myr}
E.~Tiesinga, P.~J. Mohr, D.~B. Newell and B.~N. Taylor, \emph{{CODATA
  recommended values of the fundamental physical constants: 2018*}},
  \href{https://doi.org/10.1103/RevModPhys.93.025010}{\emph{Rev. Mod. Phys.}
  {\bfseries 93} (2021) 025010}.

\bibitem{Borisyuk:2006uq}
D.~Borisyuk and A.~Kobushkin, \emph{{Two-photon exchange at low $Q^2$}},
  \href{https://doi.org/10.1103/PhysRevC.75.038202}{\emph{Phys. Rev. C}
  {\bfseries 75} (2007) 038202}
  [\href{https://arxiv.org/abs/nucl-th/0612104}{{\ttfamily nucl-th/0612104}}].

\bibitem{Blunden:2005ew}
P.~G. Blunden, W.~Melnitchouk and J.~A. Tjon, \emph{{Two-photon exchange in
  elastic electron-nucleon scattering}},
  \href{https://doi.org/10.1103/PhysRevC.72.034612}{\emph{Phys. Rev. C}
  {\bfseries 72} (2005) 034612}
  [\href{https://arxiv.org/abs/nucl-th/0506039}{{\ttfamily nucl-th/0506039}}].

\bibitem{MUSE:2013uhu}
{\scshape MUSE} collaboration, R.~Gilman et~al., \emph{{Studying the Proton
  ''Radius'' Puzzle with $\mu p$ Elastic Scattering}},
  \href{https://arxiv.org/abs/1303.2160}{{\ttfamily 1303.2160}}.

\bibitem{Tomalak:2014dja}
O.~Tomalak and M.~Vanderhaeghen, \emph{{Two-photon exchange corrections in
  elastic muon-proton scattering}},
  \href{https://doi.org/10.1103/PhysRevD.90.013006}{\emph{Phys. Rev. D}
  {\bfseries 90} (2014) 013006}
  [\href{https://arxiv.org/abs/1405.1600}{{\ttfamily 1405.1600}}].

\bibitem{Tomalak:2018jak}
O.~Tomalak and M.~Vanderhaeghen, \emph{{Dispersion relation formalism for the
  two-photon exchange correction to elastic muon\textendash{}proton scattering:
  elastic intermediate state}},
  \href{https://doi.org/10.1140/epjc/s10052-018-5988-5}{\emph{Eur. Phys. J. C}
  {\bfseries 78} (2018) 514}
  [\href{https://arxiv.org/abs/1803.05349}{{\ttfamily 1803.05349}}].

\bibitem{Engel:2022kde}
T.~Engel, \emph{{Muon-Electron Scattering at NNLO}},  {PhD Thesis},
  {Universit\"at Z\"urich}, 9, 2022.

\bibitem{Bonciani:2021okt}
R.~Bonciani et~al., \emph{{Two-Loop Four-Fermion Scattering Amplitude in QED}},
  \href{https://doi.org/10.1103/PhysRevLett.128.022002}{\emph{Phys. Rev. Lett.}
  {\bfseries 128} (2022) 022002}
  [\href{https://arxiv.org/abs/2106.13179}{{\ttfamily 2106.13179}}].

\bibitem{Buccioni:2019sur}
F.~Buccioni, J.-N. Lang, J.~M. Lindert, P.~Maierh{\"{o}}fer, S.~Pozzorini,
  H.~Zhang et~al., \emph{{OpenLoops 2}},
  \href{https://doi.org/10.1140/epjc/s10052-019-7306-2}{\emph{Eur. Phys. J. C}
  {\bfseries 79} (2019) 866}
  [\href{https://arxiv.org/abs/1907.13071}{{\ttfamily 1907.13071}}].

\bibitem{Patel:2015tea}
H.~H. Patel, \emph{{Package-X: A Mathematica package for the analytic
  calculation of one-loop integrals}},
  \href{https://doi.org/10.1016/j.cpc.2015.08.017}{\emph{Comput. Phys. Commun.}
  {\bfseries 197} (2015) 276}
  [\href{https://arxiv.org/abs/1503.01469}{{\ttfamily 1503.01469}}].

\bibitem{Djouadi:1993ss}
A.~Djouadi and P.~Gambino, \emph{{Electroweak gauge bosons selfenergies:
  Complete QCD corrections}},
  \href{https://doi.org/10.1103/PhysRevD.49.3499}{\emph{Phys. Rev. D}
  {\bfseries 49} (1994) 3499}
  [\href{https://arxiv.org/abs/hep-ph/9309298}{{\ttfamily hep-ph/9309298}}].

\bibitem{alphaqed}
F.~Jegerlehner. \url{http://www-com.physik.hu-berlin.de/~fjeger/software.html}.

\bibitem{Heller:2021mcw}
M.~Heller, N.~Keil and M.~Vanderhaeghen, \emph{{Soft-photon radiative
  corrections to the $e^-p \to e^- pl^- l^+$ process}},
  \href{https://doi.org/10.1103/PhysRevD.104.073007}{\emph{Phys. Rev. D}
  {\bfseries 104} (2021) 073007}
  [\href{https://arxiv.org/abs/2108.02088}{{\ttfamily 2108.02088}}].

\bibitem{McMule:data}
\mcmule{} Team, ``\mcmule{} dataset.''
  \url{https://doi.org/10.5281/zenodo.6541686}.

\bibitem{Workman:2022ynf}
{\scshape Particle Data Group} collaboration, R.~L. Workman and Others,
  \emph{{Review of Particle Physics}},
  \href{https://doi.org/10.1093/ptep/ptac097}{\emph{PTEP} {\bfseries 2022}
  (2022) 083C01}.

\end{thebibliography}\endgroup

\end{document}